\documentclass[article, shortnames]{jss}

\usepackage{orcidlink,thumbpdf,lmodern}
\usepackage{amsmath}
\usepackage{subcaption}
\usepackage{float}
\usepackage{amssymb}

\usepackage{etoolbox}
\apptocmd{\thebibliography}{\raggedright}{}{}
\newcommand{\cm}{\checkmark}
\newcommand{\cx}{\texttimes}

\author{
    Prashanth S Velayudhan\\Hospital for Sick Children
    \And
    Xiaoqiao Xu\\Hospital for Sick Children
    \And
    Prajkta Kallurkar\\Hospital for Sick Children
    \AND
    Ana Patricia Balbon\\University of Toronto
    \And
    Maria T Secara\\Centre for Addiction\\and Mental Health
    \And
    Adam Taback\\University of Toronto
    \And
    Denise Sabac\\Centre for Addiction\\and Mental Health
    \AND
    Nicholas Chan\\Hospital for\\Sick Children
    \And
    Shihao Ma\\University of Toronto
    \And
    Bo Wang\\University of Toronto
    \And
    Daniel Felsky\\Centre for Addiction\\and Mental Health
    \AND
    Stephanie H Ameis\\Centre for Addiction\\and Mental Health
    \And
    Brian Cox\\University of Toronto
    \And
    Colin Hawco\\Centre for Addiction\\and Mental Health
    \And
    Lauren Erdman\\Cincinnati Children's\\Hospital Medical Center
    \AND
    Anne L Wheeler\\Hospital for Sick Children
}

\Plainauthor{
    Prashanth S Velayudhan,
    Xiaoqiao Xu,
    Prajkta Kallurkar,
    Ana Patricia Balbon,
    Maria T Secara,
    Adam Taback,
    Denise Sabac,
    Nicholas Chan,
    Shihao Ma,
    Bo Wang,
    Dan Felsky,
    Stephanie H Ameis,
    Brian Cox,
    Colin Hawco,
    Lauren Erdman,
    Anne L Wheeler
}

\title{metasnf: Meta Clustering with Similarity Network Fusion in \proglang{R}}
\Plaintitle{metasnf: Meta Clustering with Similarity Network Fusion in R}
\Shorttitle{Meta Clustering with SNF in \proglang{R}}

\Abstract{
\pkg{metasnf} is an \proglang{R} package that enables users to apply meta clustering, a method for efficiently searching a broad space of cluster solutions by clustering the solutions themselves, to clustering workflows based on similarity network fusion (SNF).
SNF is a multi-modal data integration algorithm commonly used for biomedical subtype discovery.
The package also contains functions to assist with cluster visualization, characterization, and validation.
This package can help researchers identify SNF-derived cluster solutions that are guided by context-specific utility over context-agnostic measures of quality.
}

\Plainkeywords{cluster analysis, similarity network fusion, data integration, meta clustering, R}

\Address{
    Prashanth S Velayudhan, Anne L Wheeler\\
    Neurosciences and Mental Health\\
    The Hospital for Sick Children\\
    Toronto, ON, Canada\\
    \emph{and}\\
    Department of Physiology\\
    The University of Toronto\\
    Toronto, ON, Canada\\
    E-mail: \email{prashanth.velayudhan@sickkids.ca, anne.wheeler@sickkids.ca}\\

    Xiaoqiao Xu, Prajkta Kallurkar\\
    Centre for Computational Medicine\\
    Hospital for Sick Children\\
    Toronto, ON, Canada\\

    Nicholas Chan\\
    Division of Rheumatology\\
    Hospital for Sick Children\\
    Toronto, ON, Canada\\

    Shihao Ma\\
    Vector Institute\\
    Toronto, ON, Canada\\
    \emph{and}\\
    Department of Computer Science\\
    University of Toronto\\
    Toronto, ON, Canada\\
    \emph{and}\\
    Peter Munk Cardiac Centre\\
    University Health Network\\
    Toronto, ON, Canada\\

    Bo Wang\\
    Vector Institute\\
    Toronto, ON, Canada\\
    \emph{and}\\
    Department of Computer Science\\
    University of Toronto\\
    Toronto, ON, Canada\\
    \emph{and}\\
    Peter Munk Cardiac Centre\\
    University Health Network\\
    Toronto, ON, Canada\\
    \emph{and}\\
    Department of Medical Biophysics\\
    University of Toronto\\
    Toronto, ON, Canada\\

    Denise Sabac\\
    Krembil Centre for Neuroinformatics\\
    Centre for Addiction and Mental Health\\
    Toronto, ON, Canada\\
    \emph{and}\\
    Institute of Medical Science\\
    University of Toronto\\
    Toronto, ON, Canada\\

    Daniel Felsky\\
    Krembil Centre for Neuroinformatics\\
    Centre for Addiction and Mental Health\\
    Toronto, ON, Canada\\
    \emph{and}\\
    Division of Biostatistics\\
    Dalla Lana School of Public Health\\
    University of Toronto\\
    Toronto, ON, Canada\\
    \emph{and}\\
    Department of Psychiatry\\
    University of Toronto\\
    Toronto, ON, Canada\\
    \emph{and}\\
    Institute of Medical Science\\
    University of Toronto\\
    Toronto, ON, Canada\\
    \emph{and}\\
    Department of Anthropology\\
    University of Toronto\\
    Toronto, ON, Canada\\

    Stephanie H Ameis\\
    Campbell Family Mental Health Research Institute\\
    Centre for Addiction and Mental Health\\
    Toronto, ON, Canada\\
    \emph{and}\\
    Institute of Medical Science\\
    University of Toronto\\
    Toronto, ON, Canada\\
    \emph{and}\\
    Department of Psychiatry\\
    Temerty Faculty of Medicine\\
    University of Toronto\\
    Toronto, ON, Canada\\
    \emph{and}\\
    Department of Psychiatry\\
    The Hospital for Sick Children\\
    Toronto, ON, Canada\\

    Brian Cox, Ana Patricia Balbon\\
    Department of Physiology\\
    University of Toronto\\
    Toronto, ON, Canada\\
    \emph{and}\\
    Department of Obstetrics and Gynecology\\
    University of Toronto\\
    Toronto, ON, Canada\\

    Adam Taback\\
    The Edward S. Rogers Sr. Department of Electrical and Computer Engineering\\
    University of Toronto\\
    Toronto, ON, Canada\\
    
    Maria T Secara\\
    Campbell Family Mental Health Research Institute\\
    Centre for Addiction and Mental Health\\
    Toronto, ON, Canada\\
    \emph{and}\\
    Institute of Medical Science\\
    University of Toronto\\
    Toronto, ON, Canada\\

    Colin Hawco\\
    Campbell Family Mental Health Research Institute\\
    Centre for Addiction and Mental Health\\
    Toronto, ON, Canada\\
    \emph{and}\\
    Department of Psychiatry\\
    Temerty Faculty of Medicine\\
    University of Toronto\\
    Toronto, ON, Canada\\

    Lauren Erdman\\
    Division of Gastroenterology\\
    Department of Pediatrics\\
    Cincinnati Children’s Hospital Medical Center\\
    University of Cincinnati College of Medicine\\
    Cincinnati, OH, USA\\
    \emph{and}\\
    James M. Anderson Center for Health Systems Excellence\\
    Cincinnati Children’s Hospital Medical Center\\
    Cincinnati, OH, USA\\
    \emph{and}\\
    Department of Pediatrics\\
    University of Cincinnati College of Medicine\\
    Cincinnati, OH, USA
}

\begin{document}

%% -- Introduction -------------------------------------------------------------

\section{Introduction}\label{sec:intro}
Cluster analysis is frequently used in biomedical research to identify subtypes of patients with heterogeneous diseases. 
Modern workflows often employ some form of multi-omics data integration to cluster patients across features spanning diverse sources of information with minimal information loss.
One such multi-omics data integration approach is similarity network fusion (SNF) \citep{wang_similarity_2014}, which merges similarity networks derived at the scale of individual data types into a single network meant to reflect overall similarity across all data types.
Since its development, similarity network fusion has continued to demonstrate considerable success in a variety of subtyping contexts \citep{yang_multi-omic_2023,ali_longitudinal_2023,hernandez-lorenzo_genetic-based_2024,hernandez-verdin_molecular_2023,zheng_multi-omics_2024,mendez_metabolomic-derived_2024,jiang_integrated_2024}.

Typical clustering workflows, however, often share the limitations of being guided by generic metrics of clustering quality that do not necessarily align with usefulness within the context at hand and of only sampling a relatively narrow space of possible solutions.
Particularly in scenarios with large and noisy feature spaces, there are no guarantees that the most numerically compact cluster solution is the same or even remotely qualitatively similar to the cluster solution that would be maximally useful for the user's context.
What constitutes a context-optimal cluster solution is often not well-defined, and manual expert evaluation is commonly required to determine which cluster solution out of those that have been generated would likely be the most valuable for the task at hand.
For a dataset containing a non-trivial number of observations, generating and evaluating the full space of possible cluster solutions is computationally intractable.
Practically, however, this full space is expected to contain a substantial amount of redundancy; two solutions that disagree on a relatively small number of cluster assignments would likely yield very similar insights about the data.
This redundancy is leveraged by the meta clustering procedure proposed by \cite{caruana_meta_2006} to address the issues of potential disparities between context-agnostic metrics of cluster quality and context-specific usefulness as well as the relatively narrow set of cluster solutions that most workflows survey.
Meta clustering consists of first generating a wide range of possible cluster solutions through random fluctuations in clustering hyperparameters, followed by clustering the cluster solutions themselves to a manageable number of qualitatively similar ``meta clusters'' that can be manually (or automatically) evaluated for context-specific utility more rigorously than would be practical for the full space of generated solutions.

Meta clustering can be applied to any clustering workflow.
The \pkg{metasnf} package was specifically developed to facilitate meta clustering for SNF-based clustering workflows.
To our knowledge, \pkg{metasnf} is the first software package across any language to implement meta clustering as defined by \cite{caruana_meta_2006}, as well as the first software package to facilitate scalable generation of SNF-based cluster solutions.
The functionality of \pkg{metasnf} covers generating highly configurable spaces of cluster solutions, tools for visualizing and selecting meta clusters, calculations of associations between clustering and held-out features with derived solutions, traditional resampling stability and clustering quality calculations, and validation of results in held-out samples.

The implementation of SNF in \pkg{metasnf} relies on functions from the original SNF package, \pkg{SNFtool} \citep{wang_similarity_2014}.
Other software packages we are aware of that are related to SNF include the \proglang{MATLAB} package \pkg{JSNMF} \citep{ma_jsnmf_2022} and the \proglang{R} package \pkg{abSNF} \citep{ruan_using_2019}, though neither of these packages provides similar functionality to \pkg{metasnf}.

\pkg{JSNMF} implements a method called jointly semi-orthogonal non-negative matrix factorization (JSNMF) for single-cell multi-omics data integration.
This procedure includes traditional SNF as an intermediate step, but is ultimately a distinct data integration procedure that is more suitable for direct comparison with traditional SNF and other multi-omics data integration methods.
\pkg{metasnf} is not itself a data integration method, but rather a suite of functions to complement an existing and commonly used integration method (SNF).

The package \pkg{abSNF} implements association-signal-annotation boosted similarity network fusion (ab-SNF), which is an approach that rescales features based on the p-values of their associations with a held-out feature of interest.
This method can be performed within \pkg{metasnf}, as shown in Appendix~\ref{app:absnf}.

%% -- Manuscript ---------------------------------------------------------------

\section{Overview of implemented methodology}

\subsection{Similarity network fusion}

Many clustering methods require as inputs an intermediate observation-to-observation distance or similarity matrix.
In contexts such as biomedical patient subtyping, effectively integrating large and heterogeneous data modalities associated with different scales, biases, and amounts of noise into meaningful patient similarity matrices that capture the unique and complementary signal from each of those modalities has been challenging.

Similarity network fusion has empirically been demonstrated to be a very effective and efficient method for tackling this data integration problem.
The similarity network fusion pipeline consists of first establishing patient similarity matrices for each distinct source of information being integrated, then iteratively diffusing the information in those matrices to converge toward a single matrix that captures patient similarity across all of the initial sources.

\subsubsection{Distance to similarity conversion}

The inputs to similarity network fusion are $m$ similarity matrices $\mathbf{W}^1, \mathbf{W}^2, ..., \mathbf{W}^m$ of size $N \times N$  where $N$ corresponds to the number of observations and value $\mathbf{W}^{v}_{i, j}$ stores the similarity between the $i^\text{th}$ and $j^\text{th}$ observations in the $v^{\text{th}}$ matrix.
The conversion of raw data into distance matrices can be achieved in \pkg{metasnf} by a variety of common or user-defined distance metrics, discussed in more detail in Appendix~\ref{app:distance}.
To convert those distance matrices to similarity matrices, \pkg{metasnf} makes use of the \pkg{SNFtool} function \code{affinityMatrix()}.
This function takes as inputs a distance matrix as well as two hyperparameters, $K$ and $alpha$.
For each distance value between observations $i$ and $j$, a normal distribution is generated with a mean of $0$ and a standard deviation proportional to the products of $alpha$, the distance between $i$ and $j$, and the average distances of $i$ and $j$ to their $K$ nearest neighbours.
The similarity between $i$ and $j$ is then calculated as the density of the normal distribution at the value of the distance between $i$ and $j$.
A small $alpha$, low distance between $i$ and $j$, and low distance between each of $i$ and $j$ to their $K$ nearest neighbours results in a very narrow normal distribution, which in turn exaggerates the effect of observations with low distances being assigned very high similarities and observations with high distances being assigned very low similarities.

\subsubsection{Similarity network fusion}

The mathematical details of similarity network fusion are described in further detail in the original paper \citep{wang_similarity_2014}. The similarity matrices $\mathbf{W}$ are normalized such that the diagonals are equal to 0.5 and each row sums to 1 to yield matrices $\mathbf{P}$.
A new set of ``local'' similarity matrices, $\mathbf{S}^1, \mathbf{S}^2, ..., \mathbf{S}^m$, are constructed based on each observation's $K$ nearest neighbours.
If observation $j$ is among observation $i$'s $K$ nearest neighbours, the entry $\mathbf{S}(i, j)$ will be equal to the original similarity $\mathbf{W}(i, j)$, scaled down by the sum of the similarities of observation $i$ to all of its $K$ nearest neighbours.
If observation $j$ is not among the nearest neighbours of $i$, the entry $\mathbf{S}(i, j)$ will instead equal 0.
At this stage, the matrices $\mathbf{S}$ contain similarity information for observations and their nearest neighbours, while the matrices $\mathbf{P}$ contain similarity information for all pairs of observations.
Next, the matrices are iteratively fused together, according to the rule shown in Equation~\ref{eq:snf_update}:

\begin{equation} \label{eq:snf_update}
\mathbf{P}^v_{t + 1} = \mathbf{S}^v \times \frac{\sum_{k \neq v} \mathbf{P}^{k}_t}{m - 1} \times (\mathbf{S}^v)^\top
\end{equation}

This rule updates $\mathbf{P}^v$ at time step $t$ to step $t + 1$ by propagating its local similarity information ($\mathbf{S}^v$) to the average global similarity information of all other matrices.
Eventually, the matrices $\mathbf{P}$ converge into a single matrix that captures local and global observation-to-observation similarity information from all input similarity matrices.

\subsection{Meta clustering}

The meta clustering approach simply consists of generating a diverse set of cluster solutions from data by sampling across a range of reasonable clustering hyperparameters, then clustering those cluster solutions into a manageable number of groups of solutions for users to rigorously characterize.
The primary function of the \pkg{metasnf} package is to offer users a convenient way to manage cluster generation over most tunable parameters of a typical SNF-based cluster analysis pipeline.

\section{Overview of package design}

\begin{figure}[H]
\centering
\includegraphics{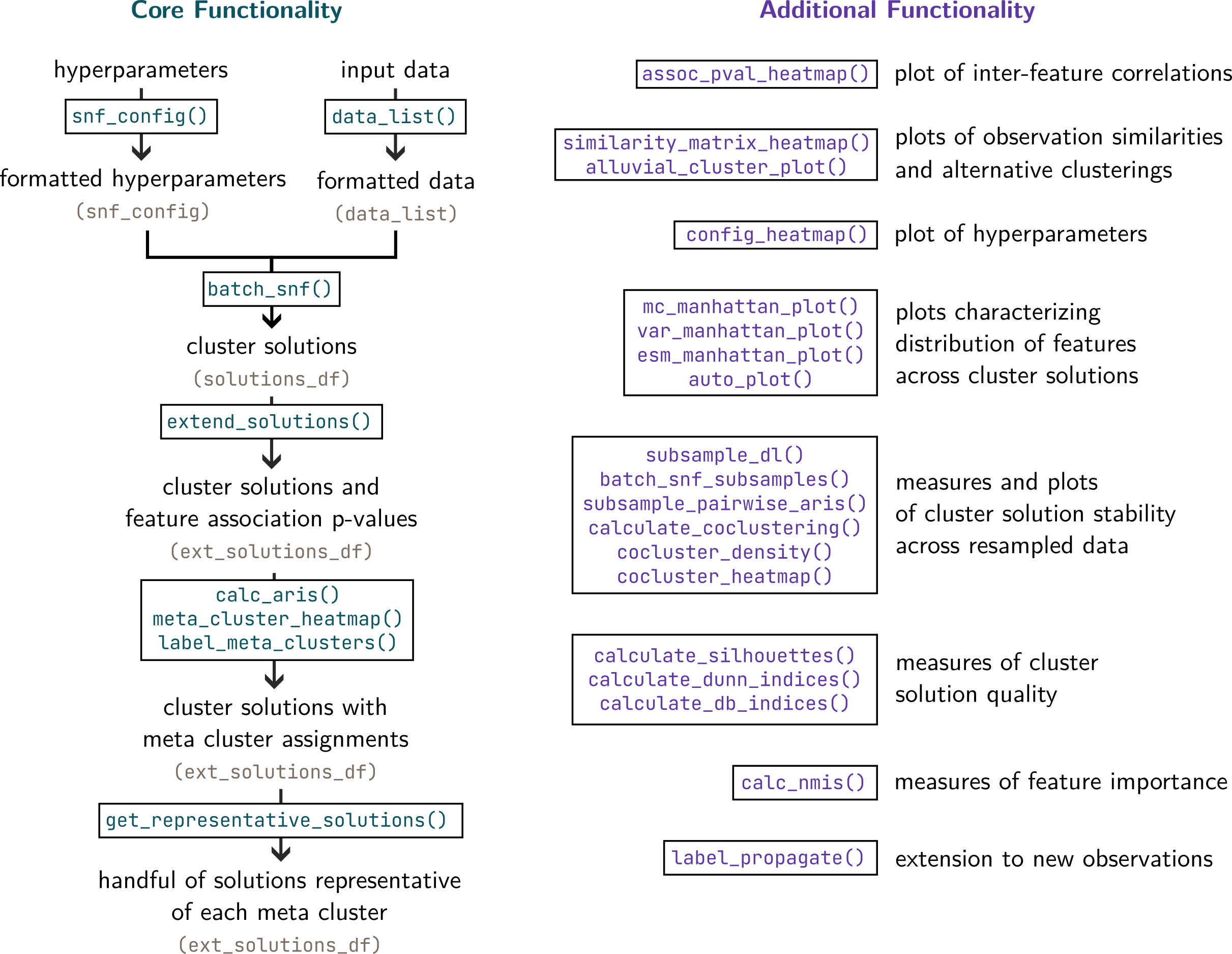}
\caption{Overview of functions in the \pkg{metasnf} package. Core functions to the \pkg{metasnf} workflow are highlighted in blue. Additional functionality for characterizing, validating, and visualizing generated results are highlighted in purple. S3 class objects are highlighted in grey.}
\label{fig:overview}
\end{figure}

The \pkg{metasnf} package makes use of \proglang{R}'s S3 methods and classes system to manage the transformation of initial data frames to well-characterized cluster solutions.
An overview of the most important functions and classes included in \pkg{metasnf} is presented in Figure~\ref{fig:overview}.
Users begin by formatting their input data for clustering as a \code{data_list} class object and by establishing the sets of hyperparameters to be used for cluster solution generation in an \code{snf_config} class object.
These objects are passed into the function \code{batch_snf()}, which yields a \code{solutions_df} (solutions data frame) class object storing information about all the final cluster assignments for all observations across all generated solutions.
The \code{extend_solutions()} function can be used to augment the \code{solutions_df} object an \code{ext_solutions_df} (extended solutions data frame) class object containing association p-values relating features and the generated cluster assignments.
Either the \code{solutions_df} or \code{ext_solutions_df} object can be used to calculate pairwise similarities between cluster solutions with the \code{calc_aris()} function, after which meta clusters (clusters of similar cluster solutions) can be visualized and delineated using the \code{meta_cluster_heatmap()} and \code{label_meta_clusters()} functions.
Representative cluster solutions from each meta cluster, as defined by the solutions that are most similar to all other solutions in their meta clusters, can be extracted using the \code{get_representative_solutions()} function into a substantially smaller set of cluster solutions for deeper characterization.

Several additional functions enable users to further characterize each step of the pipeline.
Initial associations between features prior to clustering can be examined with the \newline \code{assoc_pval_heatmap()} function.
The space of hyperparameters used for clustering can be visualized alongside the resulting cluster solutions by the \code{config_heatmap()} function.
Details of the final SNF-generated similarity matrices or variations on how they can be clustered can be visualized using the \code{similarity_matrix_heatmap()} and \code{alluvial_cluster_plot()} functions.
A variety of Manhattan plots, \code{mc_manhattan_plot()}, \code{var_manhattan_plot()}, and \code{esm_manhattan_plot()} provide different snapshots of the magnitude of feature separation across generated cluster solutions. 
The \code{auto_plot()} function produces detailed \pkg{ggplot2} \citep{ggplot} objects showing precisely how features are distributed across different cluster solutions.
Several coclustering and stability related functions exist to characterize robustness of generated solutions to data resampling.
Functions \code{calculate_silhouettes()}, \code{calculate_dunn_indices()}, and \code{calculate_db_indices()} offer convenient wrappers for calculating common metrics of clustering quality.
The \code{calc_nmis()} function calculates normalized mutual information-based measures of feature importances for constructing each cluster solution.
The \code{label_propagate()} function provides one approach to validate cluster solution generalizability by mapping results onto new, unclustered observations.

Collectively, these functions enable users to efficiently assess a broad range of possible cluster solutions that can emerge from a dataset, identify cluster solutions of maximum utility to their contexts, and deeply characterize the most important characteristics of those cluster solutions to better understand their data.

\section{Installation and set-up}\label{sec:install}

\pkg{metasnf} can be installed from the Comprehensive R Archive Network (CRAN).
The minimum requirement for installing \pkg{metasnf} is \proglang{R} version 4.1.0.
Some of the non-essential functionality of \pkg{metasnf} presented in this manuscript relies on optional dependencies.
To install \pkg{metasnf} alongside these optional dependencies, users must set the \code{dependencies} parameter value to \code{TRUE} during installation.
\begin{CodeChunk}
\begin{CodeInput}
R> install.packages("metasnf", dependencies = TRUE)
\end{CodeInput}
\end{CodeChunk}
Some visualization functions rely on the optional \proglang{R} package dependencies \pkg{ComplexHeatmap} \citep{gu_complex_2022} and \pkg{InteractiveComplexHeatmap} \citep{gu_make_2022}, which are not available on CRAN and are thus not included in the previous installation command.
These packages can be installed through the \pkg{BiocManager} package \citep{biocmanager} which is available on CRAN.
\begin{CodeChunk}
\begin{CodeInput}
R> library("BiocManager")
R> install("ComplexHeatmap")
R> install("InteractiveComplexHeatmap")
\end{CodeInput}
\end{CodeChunk}
Some of the functionality of \pkg{metasnf} is provided by directly wrapping around functions from the original \pkg{SNFtool} package to ensure within-machine reproducibility for users replicating \pkg{SNFtool}-derived results in \pkg{metasnf}.
It is important for users of \pkg{SNFtool} or \pkg{metasnf} to note that the implementation of spectral clustering provided by \pkg{SNFtool} can involve highly precise floating point calculations that in specific circumstances can propagate into distinct cluster solutions calculated from the same set of inputs across different machines.
The consensus clustering-like approach of meta clustering is an effective guard against selecting unstable cluster solutions that dependent on floating point error propagations and we do not expect the differences between machines to meaningfully change the results of the same \pkg{metasnf}-based analyses.
For purposes of exact reproducibility, we recommend users make note of their central processing unit model, Basic Linear Algebra Subprograms (BLAS) \citep{lawson1979basic} back end and number of threads used, and Linear Algebra Package (LAPACK) \citep{anderson1999lapack} version.

\section{Package overview}

\pkg{metasnf}'s functionality focuses on three core steps:

\begin{enumerate}
    \item Generation of clustering hyperparameters.
    \item Generation of cluster solutions.
    \item Selection of top cluster solution(s).
\end{enumerate}

The hyperparameters that determine how initial input data will be processed into a space of cluster solutions is stored in an object called the SNF config.
Features used to generate clusters and their metadata are structured as a nested list called the data list.
\pkg{metasnf}'s main function, \code{batch_snf}, accepts an SNF config and a data list to return a solutions data frame, a data frame that includes the corresponding clustering results for each row.
From the solutions data frame, a variety of approaches including, but not limited to, meta clustering can be used to assist users in identifying a top solution that best suits their needs.

The following example using \pkg{metasnf} demonstrates how only a few lines of code are required to generate multiple cluster solutions from a set of multi-modal data frames (provided with package installation):
\begin{CodeChunk}
\begin{CodeInput}
R> library("metasnf")
R> dl <- data_list(
+    list(subc_v, "subcortical_volume", "neuroimaging", "continuous"),
+    list(pubertal, "pubertal_status", "demographics", "continuous"),
+    uid = "unique_id")
R> config <- snf_config(dl, n_solutions = 5, max_k = 40)
R> sol_df <- batch_snf(dl, config)
\end{CodeInput}
\end{CodeChunk}

\section{Data set-up}

\subsection{Formatting raw data}

Users should load their data into the R environment in the following format:

\begin{enumerate}
    \item Stored as one or more \proglang{R} data frame class objects.
    \item In ``tidy'' \citep{tidy-data} format.
    \begin{itemize}
        \item Rows should correspond to observations to cluster.
        \item Each observation to cluster should only occur in one row per data frame.
        \item Columns should correspond to features to use for clustering or for evaluation as an out-of-model measure.
    \end{itemize}
    \item No missing values\footnote{\pkg{SNF} and by extension \pkg{metasnf} do not support analysis of incomplete data. \pkg{metasnf} can, however, be used to examine how variations in imputation strategy influence the overall structure of generated cluster solutions. This functionality is discussed further in Appendix~\ref{app:imputation}.}.
    \item One column per data frame that contains a unique identifier (UID) or key for each observation.
    \item Feature names, apart from the UID column, must be unique across all data (not just unique within their respective data frames).
\end{enumerate}

\pkg{metasnf} provides toy \proglang{R} data frames that can be inspected for reference.
These toy data frames are based on the structure of real data from the Adolescent Brain Cognitive Development Study \citep{jernigan_adolescent_2018}. An example provided mock data frame is \code{cort_t}.
\begin{CodeChunk}
\begin{CodeInput}
R> library("metasnf")
R> dim(cort_t)
\end{CodeInput}
\begin{CodeOutput}
188 152
\end{CodeOutput}
\begin{CodeInput}
R> str(cort_t[1:4, 1:5])
\end{CodeInput}
\begin{CodeOutput}
tibble [4 × 5] (S3: tbl_df/tbl/data.frame)
unique_id: chr [1:4] "NDAR_INV0567T2Y9" "NDAR_INV0GLZNC2W" ...
mrisdp_1 : num [1:4] 2.6 2.62 2.62 2.6
mrisdp_2 : num [1:4] 2.49 2.85 2.29 2.67
mrisdp_3 : num [1:4] 2.8 2.78 2.53 2.68
mrisdp_4 : num [1:4] 2.95 2.85 2.96 2.94
\end{CodeOutput}
\end{CodeChunk}
The data frame contains one UID column, \code{unique_id}, and 151 features that correspond to cortical thickness values of various parts of the brain.
Data is present for 188 subjects.
Another example data frame is ordinal household income data stored in \code{income}.
275 subjects are present in \code{income}, but only 245 of them have complete data.
In the next step of the \pkg{metasnf} pipeline, raw data will be formatted into a nested list structure called the data list, at which point incomplete data will automatically be dropped.
To avoid any surprises in the remaining number of observations, we recommend handling missing data prior to data list generation.
The \code{get_complete_uids()} helper function will reveal which observations have complete data across a list of data frames.
That information can be used to filter down each data frame to only those subjects who have complete data across all the data frames that will be used for clustering:
\begin{CodeChunk}
\begin{CodeInput}
R> all_dfs <- list(anxiety, depress, cort_t, cort_sa, subc_v, income,
+    pubertal)
R> complete_uids <- get_complete_uids(all_dfs, uid = "unique_id")
R> head(complete_uids)
\end{CodeInput}
\begin{CodeOutput}
[1] "NDAR_INV0567T2Y9" "NDAR_INV0J4PYA5F" "NDAR_INV10OMKVLE"
[4] "NDAR_INV15FPCW4O" "NDAR_INV19NB4RJK" "NDAR_INV1HLGR738"
\end{CodeOutput}
\begin{CodeInput}
R> length(complete_uids)
\end{CodeInput}
\begin{CodeOutput}
[1] 87
\end{CodeOutput}
\begin{CodeInput}
R> anxiety_df <- dplyr::filter(anxiety, unique_id %in% complete_uids)
R> depress_df <- dplyr::filter(depress, unique_id %in% complete_uids)
R> cort_t_df <- dplyr::filter(cort_t, unique_id %in% complete_uids)
R> cort_sa_df <- dplyr::filter(cort_sa, unique_id %in% complete_uids)
R> subc_v_df <- dplyr::filter(subc_v, unique_id %in% complete_uids)
R> income_df <- dplyr::filter(income, unique_id %in% complete_uids)
R> pubertal_df <- dplyr::filter(pubertal, unique_id %in% complete_uids)
\end{CodeInput}
\end{CodeChunk}

Note that \code{get_complete_uids()} only returns a vector of UIDs that have complete data; it does not do any filtering on its own.

\subsection{Creating a data list} \label{sec:data_list} 

The \code{data_list} is the primary structure in \pkg{metasnf} for storing data.
The code below creates a \code{data_list} object from 5 of our data frames.
\begin{CodeChunk}
\begin{CodeInput}
R> dl_1 <- data_list(
+    list(cort_t_df, "cortical_thickness", "neuroimaging", "continuous"),
+    list(cort_sa_df, "cortical_sa", "neuroimaging", "continuous"),
+    list(subc_v_df, "subcortical_volume", "neuroimaging", "continuous"),
+    list(income_df, "household_income", "demographics", "continuous"),
+    list(pubertal_df, "pubertal_status", "demographics", "continuous"),
+    uid = "unique_id")
\end{CodeInput}
\end{CodeChunk}
The function \code{data_list()} accepts an arbitrary number of lists, which themselves should be formatted as a list of the following four objects:
\begin{enumerate}
    \item A data frame.
    \item A name for the data frame (for user-reference).
    \item A domain of information that the data frame is associated with. Specifying domains allows users to indicate that distinct data frames are in some way related to the same broader source of information. More information on this is provided in Section~\ref{sec:partition}.
    \item The type of feature stored in the data frame. Valid options include \code{"continuous"}, \code{"discrete"}, \code{"ordinal"}, \code{"categorical"}, or \code{"mixed"} for data frames that contain more than one type of feature.
\end{enumerate}
The \code{data_list} can also be generated programmatically when dealing with a large set of data frames or with named nested list components for increased code clarity.
These approaches are shown in Appendix~\ref{app:datalist}.

A summary of the created data list can be viewed with the \code{summary()} function:
\begin{CodeChunk}
\begin{CodeInput}
R> summary(dl_1)
\end{CodeInput}
\end{CodeChunk}
\begin{CodeChunk}
\begin{CodeOutput}
                name       type       domain length width
1 cortical_thickness continuous neuroimaging     87   151
2        cortical_sa continuous neuroimaging     87   151
3 subcortical_volume continuous neuroimaging     87    30
4   household_income continuous demographics     87     1
5    pubertal_status continuous demographics     87     1
\end{CodeOutput}
\end{CodeChunk}
The summary outlines the 5 components of our data list, which all share the same 87 subjects and range from 1 to 151 features.

\subsection{Partitioning features into components for a data list} \label{sec:partition}

Users must decide how to organize the initial feature space into distinct data types, represented in code as the individual data frames supplied during data list creation.
These data types correspond to the individual data frames that are fused together in a traditional \pkg{SNFtool} run.
The specific way in which features should be organized into data types is somewhat arbitrary and ultimately a reflection of the user's perception of the level of granularity among the features that is relevant to the clustering problem at hand.
Though a set of features may have arbitrarily many layers of hierarchy to them, \pkg{metasnf} is only capable of accounting for two levels: highly similar features belong to the same data type and highly similar data types belong to the same data domain.
It may be sensible to keep multiple features related to the volumes of different subcortical brain regions in one data type, features related to the surface areas of different cortical brain regions in another data type, and features related to the volumes of different heart regions in yet another data type.
Both of the brain-related data types may belong to a neuroimaging domain, while the heart volume data type could belong to a cardiac imaging domain.
In another dataset, where a single volume feature is provided for a number of organs in the body, it may be sensible to store all those features into a single organ size data type.
In situations where feature organizing is unclear, it is recommended that users lean towards organizing their features into data types with a finer level of granularity. 

\subsection{Target data lists}

\pkg{metasnf} offers functionality to recognize a subset of features as being particularly important.
Users can store these features in another data list referred to as the target data list, .
Placing these features in a target data list makes it easier to highlight them in some downstream visualizations and guide the selection of top meta clusters and cluster solutions that are likely to be most relevant to the user's context.
It is up to users to determine if they wish to have those target features be a part of the input feature space that is used for clustering or to serve as out-of-model measures for clustering assessment.
In the code outlined in this article, we define a mock target list with features that are not a part of the original input space.
\begin{CodeChunk}
\begin{CodeInput}
R> target_dl_1 <- data_list(
+    list(anxiety_df, "anxiety", "behaviour", "ordinal"),
+    list(depress_df, "depressed", "behaviour", "ordinal"),
+    uid = "unique_id")
R> summary(target_dl_1)
\end{CodeInput}
\begin{CodeOutput}
       name    type    domain length width
1   anxiety ordinal behaviour     87     1
2 depressed ordinal behaviour     87     1
\end{CodeOutput}
\end{CodeChunk}

\section{Defining hyperparameters for clustering} \label{sec:settings}

\subsection{An overview of the SNF config}

\pkg{metasnf} offers a wide range of options to configure how cluster solutions are generated from the initial data.
By increasing the diversity and size of generated cluster solutions, users will have a better chance of finding a robust meta cluster of solutions that are maximally useful for their purposes.
A typical, complete SNF pipeline spans the following steps:
\begin{enumerate}
    \item Feature selection.
    \item Organization of features into data types.
    \item Normalization, standardization, or other rescalings of the features.
    \item Calculation of distance matrices from data types.
    \item Conversion of distance matrices to similarity (affinity) matrices.
    \item Fusion of similarity matrices by SNF.
    \item Division of the final fused matrix into a cluster solution.
\end{enumerate}
The SNF config is a \pkg{metasnf} object made up of four pieces that collectively control all parts of that pipeline:

\begin{enumerate}
    \item The settings data frame (class \code{settings_df}, extending class \code{data.frame}), which contains information about SNF-specific hyperparameters (step 4), which distance and clustering functions will be used (steps 2 and 5), and if any data frames of the data list will be excluded on a particular run (step 1). Each row of the data frame corresponds to a complete set of settings that can yield a single cluster solution from the data list.
    \item The distance functions list (class \code{dist_fns_list}, extending class \code{list}), which stores the actual distance functions that are referenced in the settings data frame (step 2).
    \item The clustering functions list (class \code{clust_fns_list}, extending class \code{list}), which similarly stores clustering functions (step 5).
    \item The weights matrix (class \code{weights_matrix}, extending classes \code{matrix} and \code{array}), which contains feature weights to account for during the data to distance matrix conversion step (step 2).
\end{enumerate}

\begin{CodeChunk}
\begin{CodeInput}
R> set.seed(42)
R> sc_1 <- snf_config(dl = dl_1, n_solutions = 20, min_k = 20, max_k = 50)
R> sc_1
\end{CodeInput}
\begin{CodeOutput}
Settings Data Frame:
                        1    2    3    4    5    6    7    8    9   10
SNF hyperparameters:
alpha                 0.5  0.4  0.3  0.3  0.5  0.4  0.7  0.8  0.3  0.6
k                      29   26   44   43   29   26   36   21   29   35
t                      20   20   20   20   20   20   20   20   20   20
SNF scheme:
                        2    1    2    1    2    2    2    3    1    3
Clustering functions:
                        1    1    2    1    2    1    1    2    1    1
Distance functions:
CNT                     1    1    1    1    1    1    1    1    1    1
DSC                     1    1    1    1    1    1    1    1    1    1
ORD                     1    1    1    1    1    1    1    1    1    1
CAT                     1    1    1    1    1    1    1    1    1    1
MIX                     1    1    1    1    1    1    1    1    1    1
Component dropout:
cortical_thickness      1    1    1    1    1    1    1    1    1    1
cortical_sa             0    1    0    1    1    1    1    1    1    1
subcortical_volume      1    1    0    0    1    1    1    1    1    1
household_income        0    1    1    1    1    1    1    0    1    0
pubertal_status         1    1    1    1    1    1    1    1    1    0
…and settings defined to create 10 more cluster solutions.
Distance Functions List:
Continuous (1):
[1] euclidean_distance
Discrete (1):
[1] euclidean_distance
Ordinal (1):
[1] euclidean_distance
Categorical (1):
[1] gower_distance
Mixed (1):
[1] gower_distance
Clustering Functions List:
[1] spectral_eigen
[2] spectral_rot
Weights Matrix:
Weights defined for 20 cluster solutions.
mrisdp_1 1, 1, 1, 1, 1, 1, 1, 1, 1, 1, 1, 1, 1, …
mrisdp_2 1, 1, 1, 1, 1, 1, 1, 1, 1, 1, 1, 1, 1, …
mrisdp_3 1, 1, 1, 1, 1, 1, 1, 1, 1, 1, 1, 1, 1, …
mrisdp_4 1, 1, 1, 1, 1, 1, 1, 1, 1, 1, 1, 1, 1, …
mrisdp_5 1, 1, 1, 1, 1, 1, 1, 1, 1, 1, 1, 1, 1, …
…and 329 more features.
\end{CodeOutput}
\end{CodeChunk}

\subsubsection{The settings data frame}

Simple options are primarily stored in the settings data frame.
The structure of the settings data frame can be viewed more closely by converting it into a regular data frame:
\begin{CodeChunk}
\begin{CodeInput}
R> as.data.frame(sc_1[["settings_df"]])[1, ]
\end{CodeInput}
\begin{CodeOutput}
solution alpha  k  t snf_scheme clust_alg cnt_dist dsc_dist ord_dist
       1   0.5 29 20          2         1        1        1        1
cat_dist mix_dist inc_cortical_thickness inc_cortical_sa
       1        1                      1               0
inc_subcortical_volume inc_household_income inc_pubertal_status
                     1                    0                   1
\end{CodeOutput}
\end{CodeChunk}
The resulting columns are:
\begin{itemize}
    \item \code{solution}: A label to keep track of each generated cluster solution.
    \item \code{alpha}: The alpha (also referred to as sigma or eta in the original SNF paper) hyperparameter in SNF. This hyperparameter plays a role in converting distance matrices into similarity matrices. The process by which \code{SNFtool::affinityMatrix()} does this conversion essentially involves plugging the distance value as the x-coordinate of a normal distribution and pulling out the density at that point as the similarity. The thickness of the normal distribution is regulated by alpha, where a larger alpha leads to a broader normal distribution and a greater sensitivity to discriminating distances.
    \item \code{k}: The nearest neighbours hyperparameter used in the distance to similarity matrix conversion as well as in similarity network fusion. In the distance matrix to similarity matrix conversion (\code{SNFtool::affinityMatrix()}), \code{k} controls how many nearest neighbours to consider when calculating how similar each observation is to its nearest neighbours. The closer an observation is to its nearest neighbours, the broader the normal distribution that is used for the distance to similarity conversion. Within the similarity network fusion step (\code{SNFtool::SNF()}), \code{k} controls how intensely all the matrices should be sparsified before information is passed between them. With a very small \code{k}, say, \code{k = 1}, all the values in all the matrices will be reduced to 0 with the exception of one value between each observation and that observation's most similar neighbour.
    \item \code{t}: The T (number of iterations) hyperparameter used in SNF. A larger \code{t} results in more rounds of information passing between similarity matrices. SNF eventually converges, so overshooting this value offers no benefit but undershooting can yield inaccurate results. The original SNF developers recommend leaving this value at 20.
    \item \code{snf_scheme}: Which of three preprocessing approaches will be used to convert the initial data frames into a final fused network. Variation in this setting is primarily done to increase the diversity of the generated cluster solutions.
    \begin{enumerate}
        \item Individual: Each individual input data frame provided in the data list is converted to its own distance matrix and subsequently its own similarity matrix. Then, all similarity matrices are combined into the final network by SNF. This approach matches the traditional \pkg{SNFtool} pipeline. One important characteristic of this approach is that the integration is biased towards information sources (domains) that have disproportionately higher representation among all data frames. For example, applying the individual scheme to a combination of 5 demographic data frames and 10 neuroimaging data frames will yield an integration that is biased towards the neuroimaging data.
        \item Two-step: Input data frames are combined within user-specified data domains by one round of SNF and then combined across domains by a second round of SNF. This scheme is intended to reduce some of the weighting issues in the individual scheme. As applied to the individual example, the final integration would involve combining a single demographic similarity matrix with a single neuroimaging similarity matrix, resulting in a more balanced integration at the domain level.
        \item Domain concatenation: Input data frames are first combined within domains by simple concatenation. Domain-based data frames are then combined into a single fused network by SNF. This approach can lead to the greatest information loss during the distance matrix calculation step, but offers additional diversity to the space of generated cluster solutions.
    \end{enumerate}
    \item Columns ending in \code{dist}: Which distance metric is being used for the various types of features (continuous, discrete, ordinal, categorical, and mixed). Users can supply their own distance metrics to apply.
    \begin{enumerate}
        \item Euclidean distance is the default used for continuous, discrete, and ordinal features.
        \item Gower distance \citep{gower_general_1971} from the R package \pkg{cluster} \citep{maechler_2023} is the default used for categorical and mixed features.
    \end{enumerate}
    \item \code{clust_alg}: Which clustering algorithm will be applied to the final fused network. Users can supply their own clustering algorithms to apply.
    \begin{enumerate}
        \item Spectral clustering with the number of clusters set by the eigen-gap heuristic.
        \item Spectral clustering with the number of clusters set by the rotation cost heuristic.
    \end{enumerate}
    \item Columns starting with \code{inc_}: Whether or not the corresponding data frame will be included or dropped from a particular SNF run.
\end{itemize}
Through variation in these columns, users can define distinct SNF pipelines that give rise to a broad space of cluster solutions.

\subsubsection{The distance and clustering functions lists}

While the settings data frame stores information on which of the available distance and clustering functions should be used for each type of data, the distance and clustering functions lists store the functions themselves.
The structure of either of these objects can be viewed more closely by converting them into a regular list.

\begin{CodeChunk}
\begin{CodeInput}
R> as.list(sc_1[["dist_fns_list"]])[[1]]
\end{CodeInput}
\begin{CodeOutput}
euclidean_distance
function (df, weights_row)
{
    weights <- diag(weights_row, nrow = length(weights_row))
    weighted_df <- as.matrix(df) %*% weights
    distance_matrix <- as.matrix(
        stats::dist(weighted_df, method = "euclidean")
    )
    return(distance_matrix)
}
\end{CodeOutput}
\end{CodeChunk}
The distance functions list is structured as a list of lists for each type of data (continuous, discrete, ordinal, categorical, and mixed), each of which stores an arbitrary number of functions converting data frames into distance matrices that can be referenced by the settings data frame.
The clustering functions list is simply a regular list of functions that can convert a final similarity matrix into a cluster solution (i.e., a numeric vector of cluster assignments).
More information on customizing the distance and clustering functions used in \pkg{metasnf} is presented in Appendices~\ref{app:distance} and \ref{app:cluster}.

\subsubsection{The weights matrix}

The weights matrix is a simple matrix containing information on how features should be weighted prior to distance matrix calculations.
Weights for each feature are stored along the columns of the matrix.
Each row contains a full set of feature weights used for generating a single cluster solution.
\begin{CodeChunk}
\begin{CodeInput}
as.matrix(sc_1[["weights_matrix"]])[1:5, 1:5]
\end{CodeInput}
\begin{CodeOutput}
     mrisdp_1 mrisdp_2 mrisdp_3 mrisdp_4 mrisdp_5
[1,]        1        1        1        1        1
[2,]        1        1        1        1        1
[3,]        1        1        1        1        1
[4,]        1        1        1        1        1
[5,]        1        1        1        1        1
\end{CodeOutput}
\begin{CodeInput}
dim(sc_1[["weights_matrix"]])
\end{CodeInput}
\begin{CodeOutput}
20 334
\end{CodeOutput}
\end{CodeChunk}
By default, all feature weights for all generated cluster solutions are set to \code{1} (i.e., unweighted).
The space of feature weights can also be randomly generated from a uniform or exponential distribution by specifying the \code{weights_fill} parameter in \code{snf_config()}.

\subsection{Constructing an SNF config}

When not specifying any parameters beyond the number of cluster solutions to be generated, calling \code{snf_config()} will result in an \code{snf_config} class object with values randomly varied across sensible default ranges.
This section outlines how the variation in the different tunable parameters of an SNF config can be controlled.

\subsubsection{Alpha, k, and t}

Hyperparameters alpha, k, and t can be controlled either by providing a vector of possible values or a range to sample from.
\begin{CodeChunk}
\begin{CodeInput}
R> sc_2 <- snf_config(dl_1, n_solutions = 100, min_k = 10, max_k = 60,
+    min_alpha = 0.3, max_alpha = 0.8)
R> sc_3 <- snf_config(dl_1, n_solutions = 20, k_values = c(10, 25, 50),
+    alpha_values = c(0.4, 0.8))
\end{CodeInput}
\end{CodeChunk}

\subsubsection{Data frame exclusion}

By default, the settings data frame generated during the call to \code{snf_config()} will randomly assign some data frames to be excluded during the generation of each cluster solution.
The number of data frames retained (not dropped) during the generation of each cluster solution is determined by sampling from an exponentially decaying probability distribution.
Consequently, the probability of dropping a small number of data frames is much greater than the probability of dropping a large number of data frames.
The probability distribution used to determine the number of dropped data frames can be adjusted by changing the \code{dropout_dist} parameter in \code{snf_config()} from the default value of \code{"exponential"} to either \code{"uniform"} (which will result in a much higher number of data frames being dropped on average) or "none" (which will result in no data frames being dropped).

\begin{CodeChunk}
\begin{CodeInput}
R> set.seed(42)
R> sc_4 <- snf_config(dl_1, n_solutions = 20, dropout_dist = "uniform")
R> sc_4[["settings_df"]]
\end{CodeInput}
\begin{CodeOutput}
                        1    2    3    4    5    6    7    8    9   10
SNF hyperparameters:
alpha                 0.3  0.7  0.7  0.5  0.6  0.7  0.7  0.3  0.8  0.8
k                      56   14   78   30   18   64   90   64   15   15
t                      20   20   20   20   20   20   20   20   20   20
SNF scheme:
                        2    3    2    2    1    1    2    2    1    2
Clustering functions:
                        1    2    2    1    1    1    1    2    1    2
Distance functions:
CNT                     1    1    1    1    1    1    1    1    1    1
DSC                     1    1    1    1    1    1    1    1    1    1
ORD                     1    1    1    1    1    1    1    1    1    1
CAT                     1    1    1    1    1    1    1    1    1    1
MIX                     1    1    1    1    1    1    1    1    1    1
Component dropout:
cortical_thickness      1    1    1    1    1    0    1    1    0    1
cortical_sa             1    1    1    1    1    0    0    1    1    1
subcortical_volume      1    1    1    1    1    1    1    1    1    1
household_income        1    0    1    1    1    0    1    0    1    1
pubertal_status         1    1    0    1    1    0    1    1    0    1
…and settings defined to create 10 more cluster solutions.
\end{CodeOutput}
\end{CodeChunk}

The bounds on the number of data frames that are dropped can also be controlled using the \code{min_removed_inputs} and \code{max_removed_inputs} parameters.

\subsubsection{Piecewise construction}

Users can also explore different sets of solution spaces by constructing SNF configs piecewise using the \code{merge()} function.
The example below combines an SNF config covering 25 solutions generated with \code{k = 50} and another 25 solutions generated with \code{k = 80}.
The two SNF configs being merged must have been generated from the same data list.

\begin{CodeChunk}
\begin{CodeInput}
R> sc_5 <- merge(snf_config(dl_1, n_solutions = 25, k_values = 50),
+    snf_config(dl_1, n_solutions = 25, k_values = 50))
\end{CodeInput}
\end{CodeChunk}

\section{Generating cluster solutions}

\subsection{Batch SNF}

Once the SNF config and data list have been defined, the \code{batch_snf()} function can be used to generate cluster solutions stored as a \code{solutions_df} class object.
\begin{CodeChunk}
\begin{CodeInput}
R> set.seed(42)
R> sc_6 <- snf_config(dl_1, n_solutions = 20, min_k = 20, max_k = 50)
R> sol_df_1 <- batch_snf(dl_1, sc_6)
R> print(sol_df_1, n = 2)
\end{CodeInput}
\begin{CodeOutput}
20 cluster solutions of 87 observations:
solution nclust mc uid_NDAR_INV0567T2Y9 uid_NDAR_INV0J4PYA5F
       1      5  .                    5                    2
       2      3  .                    3                    3
18 solutions and 85 observations not shown.
\end{CodeOutput}
\end{CodeChunk}
\code{batch_snf()} can make use of parallel processing in multicore systems by adjusting the \code{processes} parameter.
This parameter can either be set to \code{"max"} to use as many threads as possible, or can be set to a specific integer value below the total available threads on the system.
Parallelization in \pkg{metasnf} is achieved using the \pkg{future} package ecosystem \citep{bengtsson_future}.
As there is some overhead cost of parallel processing, it is recommended that users try benchmarking performance trade-offs between \code{processes = 1} and \code{processes = "max"} at a small scale first to determine if parallelization is worth enabling on their systems.

Progress of the function can be monitored regardless of parallelization using the \code{with_progress()} function from the \pkg{progressr} package \citep{bengtsson_progressr}.
\begin{CodeChunk}
\begin{CodeInput}
R> progressr::with_progress({sol_df_1 <- batch_snf(dl_1, sc_6)})
\end{CodeInput}
\end{CodeChunk}
Note that usage of \pkg{progressr} can slow down the overall time taken by \code{batch_snf()}.

\subsection{The solutions data frame}

The solutions data frame (class \code{solutions_df}) is a \pkg{metasnf} object containing the assigned clusters of all observations for each defined SNF run.
The solutions data frame contains a column to index the contained solutions (\code{solution}), a column to indicate how many clusters each solution has (\code{nclust}), which meta cluster each solution belongs to (\code{mc}; more on this in Section~\ref{sec:metacluster}), and one column for every observation indicating which cluster they were assigned for each solution (\code{uid_*}).
In the solutions data frame shown above, observation \code{NDAR_INV0567T2Y9} was assigned to cluster \code{5} in the first cluster solution (a 5-cluster solution) and to cluster \code{3} in the second cluster solution (a 3-cluster solution).

The solutions data frame stores the SNF config used to create it as an attribute accessible by the command \code{attr(sol_df, "snf_config")}.

Calling the transpose function \code{t()} on a solutions data frame object reformats the object such that each column corresponds to a cluster solution and observations are stored along the rows.
While \pkg{metasnf} contains a variety of functions to aid in the analysis and characterization of the generated cluster solutions, the transposed form may be more useful to users running analytical functions outside of \pkg{metasnf}.
\begin{CodeChunk}
\begin{CodeInput}
R> t(sol_df_1)
\end{CodeInput}
\begin{CodeOutput}
20 cluster solutions of 87 observations:
                 uid      s1    s2    s3    s4    s5    s6    s7    s8
uid_NDAR_INV0567T2Y9       5     3     1     1     1     1     1     3
uid_NDAR_INV0J4PYA5F       2     3     7     2     6     2     4     4
uid_NDAR_INV10OMKVLE       1     2     3     1     4     2     3     1
uid_NDAR_INV15FPCW4O       1     2     5     2     5     2     3     1
uid_NDAR_INV19NB4RJK       4     2     9     1     7     2     3     2
uid_NDAR_INV1HLGR738       4     2     9     1     7     2     3     2
uid_NDAR_INV1KR0EZFU       4     2     9     1     7     2     3     2
uid_NDAR_INV1L3Y9EOP       1     2     5     2     5     2     3     1
uid_NDAR_INV1TCP5GNM       1     2     3     1     4     2     3     1
uid_NDAR_INV1ZHRDJ6B       3     1     4     1     7     1     3     5
12 solutions and 77 observations not shown.
\end{CodeOutput}
\end{CodeChunk}

\section{Meta clustering}\label{sec:metacluster}

\subsection{Adjusted Rand indices}
Meta clustering enables efficient characterization of the generated cluster solutions by clustering those solutions into a manageable number of groups of solutions, referred to as meta clusters.
Pairwise similarities between cluster solutions are calculated in \pkg{metasnf} through the adjusted Rand index (ARI; \cite{hubert_comparing_1985}).
The original Rand index \citep{rand_objective_1971} is calculated as the fraction of all pairs of observations that clustered consistently (i.e., together in both solutions or apart in both solutions) between two cluster solutions.
The ARI is an adjusted-for-chance variation of the Rand index that takes on a value of 1 for two identical cluster solutions, 0 for two cluster solutions that share the same similarity as two random cluster solutions, and values below 0 for cluster solutions that are less similar to each other than random clusterings.

\subsection{Defining and visualizing meta clusters}

\begin{figure}[H]
\centering
\includegraphics[width=0.6\textwidth]{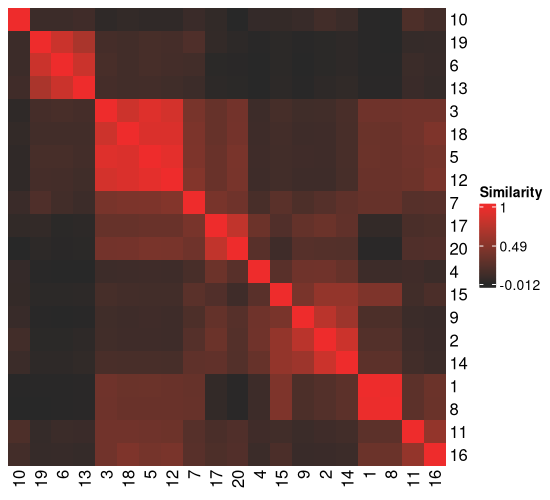}
\caption{
    Heatmap of adjusted Rand indices between 20 generated cluster solutions.
}
\label{fig:ari_hm}
\end{figure}

Adjusted Rand indices can be calculated from the solutions data frame using the \code{calc_aris()} function.
\begin{CodeChunk}
\begin{CodeInput}
R> ari_matrix_1 <- calc_aris(sol_df_1)
R> ari_matrix_1[1:3, 1:3]
\end{CodeInput}
\begin{CodeOutput}
         1         2         3
1 1.000000 0.2075450 0.3577450
2 0.207545 1.0000000 0.1154485
3 0.357745 0.1154485 1.0000000
\end{CodeOutput}
\begin{CodeInput}
R> dim(ari_matrix_1)
\end{CodeInput}
\begin{CodeOutput}
[1] 20 20
\end{CodeOutput}
\end{CodeChunk}
Complete-linkage, Euclidean distance-based hierarchical clustering is applied to the resulting matrix to obtain a sensible ordering of the generated solutions.
Alternative distance methods (of those provided by \code{stats::dist()}) and agglomerative hierarchical clustering methods (of those provided by \code{stats::hclust()}) can be specified through the parameters \code{dist_method} and \code{hclust_method}.
\begin{CodeChunk}
\begin{CodeInput}
R> meta_cluster_order_1 <- get_matrix_order(ari_matrix_1)
R> meta_cluster_order_1
\end{CodeInput}
\begin{CodeOutput}
[1] 10 19  6 13  3 18  5 12  7 17 20  4 15  9  2 14  1  8 11 16
\end{CodeOutput}
\end{CodeChunk}
The adjusted Rand index matrix and its hierarchical clustering-based order can then be visualized as the base of the meta cluster heatmap for all the generated solutions (Figure~\ref{fig:ari_hm}).
Note that the function \code{meta_cluster_heatmap()} used below is a descriptive alias for the method \code{plot.ari_matrix()}.
\begin{CodeChunk}
\begin{CodeInput}
R> ari_hm_1 <- meta_cluster_heatmap(ari_matrix_1,
+    order = meta_cluster_order_1, show_row_names = TRUE,
+    show_column_names = TRUE)
R> ari_hm_1
\end{CodeInput}
\end{CodeChunk}

Groups of cluster solutions that are highly similar form visually discernible squares in the sorted adjusted Rand index heatmap.
A vector of indices to partition these meta clusters, referred to here as the split vector, can be passed into \code{meta_cluster_heatmap()}.
This vector can be obtained by trial and error, or through interacting with the heatmap in the \code{shiny_annotator()} function.
\begin{CodeChunk}
\begin{CodeInput}
R> shiny_annotator(ari_hm_1)
\end{CodeInput}
\end{CodeChunk}
This function launches an R Shiny \citep{rshiny} application where users can identify the indices of meta cluster boundaries on an adjusted Rand index heatmap by clicking on them.
The function itself wraps around Shiny functionality provided by \pkg{InteractiveComplexHeatmap} \citep{gu_make_2022}.

\begin{figure}[H]
\centering
\includegraphics[width=0.6\textwidth]{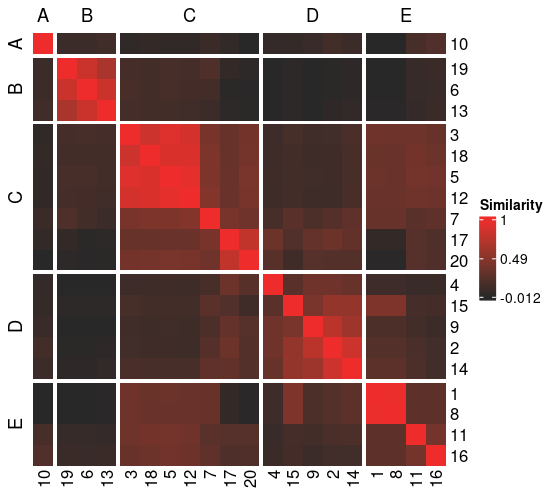}
\caption{
    Heatmap of adjusted Rand indices, partitioned into meta clusters A-E.
}
\label{fig:ari_mc_hm}
\end{figure}

Once the split vector values have been identified, the partitioned heatmap can be generated (Figure~\ref{fig:ari_mc_hm}).
\begin{CodeChunk}
\begin{CodeInput}
R> split_vec_1 <- c(2, 5, 12, 17)
R> ari_mc_hm_1 <- meta_cluster_heatmap(ari_matrix_1,
+    order = meta_cluster_order_1, split_vector = split_vec_1,
+    show_row_names = TRUE, show_column_names = TRUE)
R> ari_mc_hm_1
\end{CodeInput}
\end{CodeChunk}

Once a suitable split vector has been identified, the meta cluster column (\code{"mc"}) within the solutions data frame can be updated with the \code{label_meta_clusters()} function.
\begin{CodeChunk}
\begin{CodeInput}
R> mc_sol_df_1 <- label_meta_clusters(sol_df_1, order = meta_cluster_order_1,
+    split_vector = split_vec_1)
R> print(mc_sol_df_1, n = 2)
\end{CodeInput}
\end{CodeChunk}

\begin{CodeChunk}
\begin{CodeOutput}
20 cluster solutions of 87 observations:
solution nclust mc uid_NDAR_INV0567T2Y9 uid_NDAR_INV0J4PYA5F
       1      5 E                     5                    2
       2      3 D                     3                    3
18 solutions and 85 observations not shown.
\end{CodeOutput}
\end{CodeChunk}

\subsection{Identifying representative solutions}

Within each meta cluster, the cluster solution with the highest overall adjusted Rand indices to all other solutions in that meta cluster can be considered as being that meta cluster's most representative solution; it is the most similar solution to all other solutions in its group.
The solutions data frame can be filtered down to just those representative solutions as follows:
\begin{CodeChunk}
\begin{CodeInput}
R> rep_sol_df_1 <- get_representative_solutions(ari_matrix_1, mc_sol_df_1)
R> rep_sol_df_1
\end{CodeInput}
\end{CodeChunk}
\newpage
\begin{CodeChunk}
\begin{CodeOutput}
5 cluster solutions of 87 observations:
solution nclust mc uid_NDAR_INV0567T2Y9 uid_NDAR_INV0J4PYA5F
       1      5 E                     5                    2
       5      8 C                     1                    6
       6      2 B                     1                    2
      10      4 A                     2                    4
      14      3 D                     3                    3
85 observations not shown.
\end{CodeOutput}
\end{CodeChunk}
Now, there is only one representative row in the solutions data frame \code{rep_solutions} for each identified meta cluster.
At this point, users have completed the core steps of meta clustering with SNF.
The cluster solutions present in the \code{rep_solutions} data frame may be characterized as in a normal clustering workflow to identify a top solution based on the user's context.
The following sections outline additional functionality offered in \pkg{metasnf} to assist with the characterization, validation, and visualization of these generated cluster solutions.

\section{Evaluating cluster separation across features}

\subsection{Extending the solutions data frame}

The \code{extend_solutions()} function can be used to generate p-values related to the strength of the association between cluster solutions and individual features based on the types of those features.
The p-value comes from the linear model \code{cluster ~ feature} for continuous and discrete features, from the ordinal regression model \code{cluster ~ feature} for ordinal features, and from either the Chi-squared test (default) or Fisher's exact test (if specified) for the pair \code{cluster, feature} for categorical features.
\begin{CodeChunk}
\begin{CodeInput}
R> ext_sol_df_1 <- extend_solutions(mc_sol_df_1, target_dl = target_dl_1,
+    dl = dl_1)
R> print(ext_sol_df_1, n = 2)
\end{CodeInput}
\begin{CodeOutput}
20 cluster solutions, 87 observations, and p-values for 336 features.
Cluster assignment columns:
solution nclust mc uid_NDAR_INV0567T2Y9 uid_NDAR_INV0J4PYA5F
       1      5 E                     5                    2
       2      3 D                     3                    3
Association p-value columns:
solution mrisdp_1_pval mrisdp_2_pval mrisdp_3_pval mrisdp_4_pval
       1 4.0778e-01    8.7353e-01    1.8227e-01    6.4527e-01
       2 2.9963e-01    3.7253e-01    6.0899e-01    2.9392e-02
Summary p-value columns:
solution min_pval  mean_pval max_pval
       1 7.263e-01 7.304e-01 7.344e-01
       2 3.238e-01 4.941e-01 6.644e-01
Summaries calculated from 2 features. Use `summary_features(x)` to see them.
18 solutions and 332 features not shown.
\end{CodeOutput}
\end{CodeChunk}
Internally, the extended solutions data frame is a horizontal extension of the solutions data frame with newly added columns containing the association p-values of the features in the provided data list with all the cluster solutions.
There are additionally three summary columns, \code{min_pval}, \code{mean_pval}, \code{max_pval}, to help users quickly evaluate how well the corresponding cluster solutions were able to separate the observations based on the features in the target data list.
The only distinction between providing data through the \code{dl} parameter and the \code{target_dl} parameter is that the features in \code{dl} are not included during the calculation of the three summary measures.

\subsection{Heatmap annotations} \label{sec:hm_annotations}

\begin{figure}[H]
\centering
\includegraphics{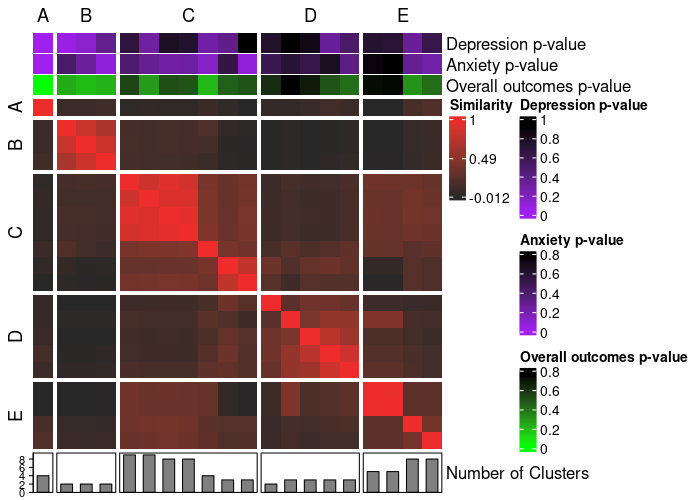}
\caption{
    Annotated heatmap of adjusted Rand indices.
}
\label{fig:annotated_ari_hm}
\end{figure}

The annotation functionality offered by the \pkg{ComplexHeatmap} package makes it easy to visualize a variety of attributes of the generated cluster solutions.
The code below demonstrates how to prepare Figure~\ref{fig:annotated_ari_hm}.
\begin{CodeChunk}
\begin{CodeInput}
R> annotated_ari_hm_1 <- meta_cluster_heatmap(x = ari_matrix_1,
+    order = meta_cluster_order_1, split_vector = split_vec_1,
+    data = ext_sol_df_1,
+    top_hm = list(
+      "Depression p-value" = "cbcl_depress_r_pval",
+      "Anxiety p-value" = "cbcl_anxiety_r_pval",
+      "Overall outcomes p-value" = "mean_pval"),
+    bottom_bar = list("Number of Clusters" = "nclust"),
+    annotation_colours = list(
+      "Depression p-value" = colour_scale(
+        ext_sol_df_1[, "cbcl_depress_r_pval"], min_colour = "purple",
+        max_colour = "black"),
+      "Anxiety p-value" = colour_scale(
+        ext_sol_df_1[, "cbcl_anxiety_r_pval"], min_colour = "purple",
+        max_colour = "black"),
+      "Overall outcomes p-value" = colour_scale(
+        ext_sol_df_1[, "mean_pval"], min_colour = "green",
+        max_colour = "black")))
R> annotated_ari_hm_1
\end{CodeInput}
\end{CodeChunk}

The \code{meta_cluster_heatmap} function is ultimately a wrapper for the \code{Heatmap()} function from \pkg{ComplexHeatmap}.
As such, guidance on additional customization options for this and any other heatmap in \pkg{metasnf} can be found within the \pkg{ComplexHeatmap} package's documentation.

\begin{figure}[H]
\centering
\includegraphics{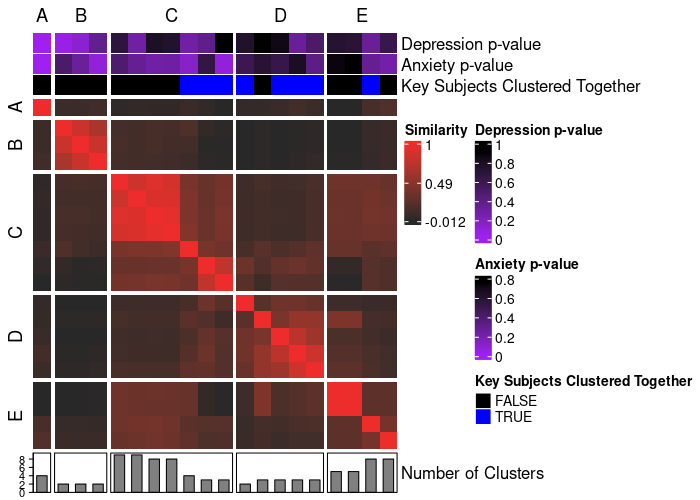}
\caption{
    Annotated heatmap of adjusted Rand indices with a manually generated annotation indicating if two specific subjects clustered together.
}
\label{fig:annotated_ari_hm2}
\end{figure}

Adding new data to visualize for heatmap annotations can be done by further extending the columns of the extended solutions data frame manually.
For example, the chunk of code below could be used to highlight cluster solutions in which two specific observations clustered together (Figure~\ref{fig:annotated_ari_hm2}).
This and related visualizations can be helpful in prioritizing deeper characterizations of the generated meta clusters.
\begin{CodeChunk}
\begin{CodeInput}
R> annotation_data_1 <- ext_sol_df_1 |>
+    as.data.frame(keep_attributes = TRUE) |>
+    dplyr::mutate(key_subjects_cluster_together = dplyr::case_when(
+      uid_NDAR_INVLF3TNDUZ == uid_NDAR_INVLDQH8ATK ~ TRUE, TRUE ~ FALSE))
R> annotated_ari_hm_2 <- meta_cluster_heatmap(
+    ari_matrix_1,
+    order = meta_cluster_order_1,
+    split_vector = split_vec_1,
+    data = annotation_data_1,
+    top_hm = list(
+      "Depression p-value" = "cbcl_depress_r_pval",
+      "Anxiety p-value" = "cbcl_anxiety_r_pval",
+      "Key Subjects Clustered Together" = "key_subjects_cluster_together"),
+    bottom_bar = list("Number of Clusters" = "nclust"),
+    annotation_colours = list(
+      "Depression p-value" = colour_scale(
+        ext_sol_df_1[, "cbcl_depress_r_pval"],
+        min_colour = "purple", max_colour = "black"),
+      "Anxiety p-value" = colour_scale(
+        ext_sol_df_1[, "cbcl_anxiety_r_pval"],
+        min_colour = "purple", max_colour = "black"),
+      "Key Subjects Clustered Together" = c(
+        "TRUE" = "blue", "FALSE" = "black")))
R> annotated_ari_hm_2
\end{CodeInput}
\end{CodeChunk}

\subsection{Meta cluster Manhattan plots} \label{sec:manhattan}

\begin{figure}[H]
\centering
\includegraphics{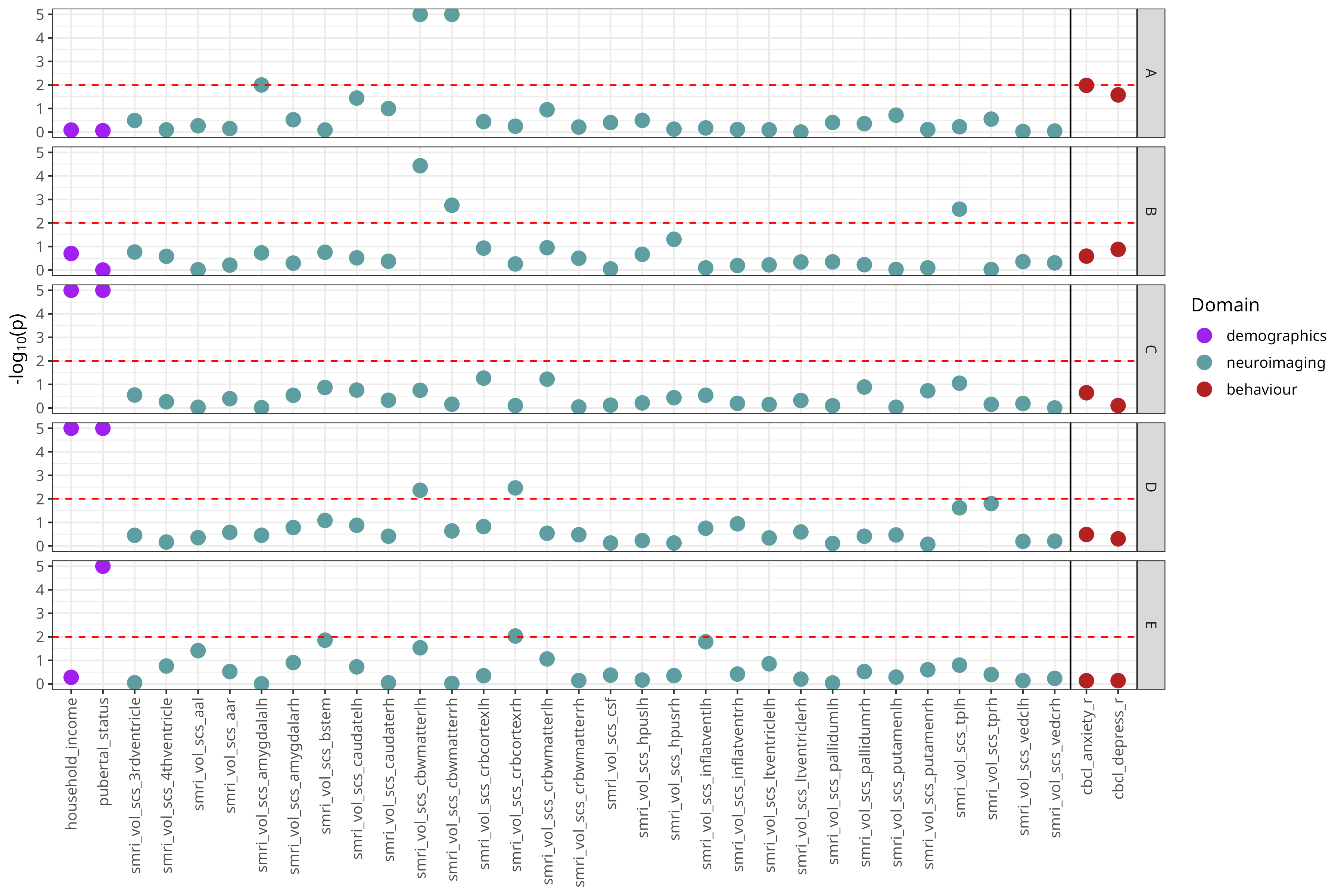}
\caption{
    Manhattan plot showing separation of all features for representative cluster solutions from each meta cluster.
    Input and target features are along the x-axis.
    A small vertical line separates the features provided through the data list (left) from those provided through the target list (right).
    Y-axis values represent $-log_{10}(\text{p-values})$ calculated for the extended solutions data frame.
    Colours of the points reflect the domains that the features belong to.
}
\label{fig:mc_manhattan2}
\end{figure}

Manhattan plots are effective tools for visualizing magnitudes of association p-values for multiple cluster solutions simultaneously.
Manhattan plots in \pkg{metasnf} are created using the \pkg{ggplot2} package \citep{ggplot}.
The Manhattan plot in Figure~\ref{fig:mc_manhattan2} shows the $-log_{10}(\text{p-value})$ of the associations between representative cluster solutions and the input and target features used to generate them.
The parameter \code{threshold = 5} is used to threshold any p-values below $10^{-5}$ to $10^{-5}$.
A large subset of neuroimaging features are removed in this example to reduce visual clutter.
\begin{CodeChunk}
\begin{CodeInput}
R> rep_sol_df_2 <- get_representative_solutions(ari_matrix_1, ext_sol_df_1)
R> rep_sol_df_2 <- dplyr::select(rep_sol_df_2, -dplyr::contains("mrisdp"))
R> palette <- c("neuroimaging" = "cadetblue", "demographics" = "purple",
+    "behaviour" = "firebrick")
R> mc_manhattan_1 <- mc_manhattan_plot(rep_sol_df_2, dl = dl_1,
+    target_dl = target_dl_1, point_size = 2, threshold = 0.05,
+    text_size = 12, domain_colours = palette)
R> mc_manhattan_1
\end{CodeInput}
\end{CodeChunk}

\subsection{P-value heatmaps}

\begin{figure}[H]
\centering
\includegraphics[width=0.5\textwidth]{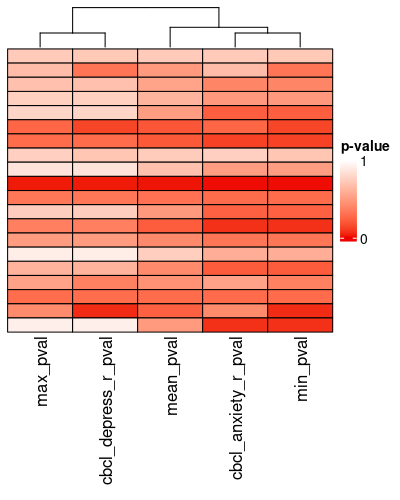}
\caption{
    Heatmap of p-values of associations between features in target list and metasnf-generated cluster solutions.
}
\label{fig:pval_hm}
\end{figure}

Similar to the Manhattan plots, p-values in the extended solutions data frame can be visualized with the \code{pval_heatmap()} function (Figure~\ref{fig:pval_hm}).
\begin{CodeChunk}
\begin{CodeInput}
R> ext_sol_df_2 <- extend_solutions(mc_sol_df_1, target_dl_1)
R> pval_heatmap(ext_sol_df_2)
\end{CodeInput}
\end{CodeChunk}

\subsection{Normalized mutual information}

Normalized mutual information (NMI), a normalized variation of mutual information \citep{shannon_mathematical_1948}, was used in the original \pkg{SNFtool} package as a method to determine the relative importance of the input features used to generate a particular cluster solution.
The approach is based on the premise that if a feature was very important in driving the generation of a particular cluster solution, generating a cluster solution from that feature alone should result in a solution that is very similar (has a high mutual information with) the original solution.
In the \pkg{SNFtool} implementation, the cluster solution based on the individual feature being assessed was assumed to be generated using squared Euclidean distance, a K hyperparameter value of 20, an alpha hyperparameter value of 0.5, and spectral clustering with the number of clusters based on the best eigen-gap value of possible solutions spanning from 2 to 5 clusters.
In contrast, the \pkg{metasnf} implementation leverages all the architectural details and hyperparameters supplied in the original \code{snf_config()} and \code{batch_snf()} call to make the solo-feature to all-feature solutions as comparable as possible.
Note that calculation of NMI values below involves spectral clustering of highly unstable cluster solutions; Section~\ref{sec:install} outlines why exact reproducibility of those values is not guaranteed across all machines.
\begin{CodeChunk}
\begin{CodeInput}
R> calc_nmis(dl = dl_1[4:5], sol_df = sol_df_1)
\end{CodeInput}
\end{CodeChunk}
\newpage
\begin{CodeChunk}
\begin{CodeOutput}
         feature         s1        s2        s3        s4        s5
household_income 0.03296816 0.2327372 0.7050127 0.2607621 0.7210830
 pubertal_status 0.95192185 0.3019173 0.6041225 0.1848260 0.5671576
        s6        s7         s8        s9        s10       s11       s12
0.01362196 0.4980339 0.03613476 0.2356433 0.01994138 0.3662702 0.7207011
0.01917553 0.3844686 0.94307926 0.2384696 0.02914684 0.4551426 0.5450593
       s13       s14       s15       s16       s17       s18        s19
0.01583377 0.2462041 0.1117085 0.3856425 0.7298814 0.6934028 0.05203014
0.04175697 0.3298108 0.5732265 0.4866952 0.1082141 0.5492709 0.02227085
       s20
1.00000000
0.05996579
\end{CodeOutput}
\end{CodeChunk}

\section{Identifying settings driving meta cluster formation}

\begin{figure}[H]
    \centering
    \begin{minipage}{0.5\textwidth}
        \centering
        \vspace{2.2cm} % Adjust the vertical offset as needed
        \includegraphics{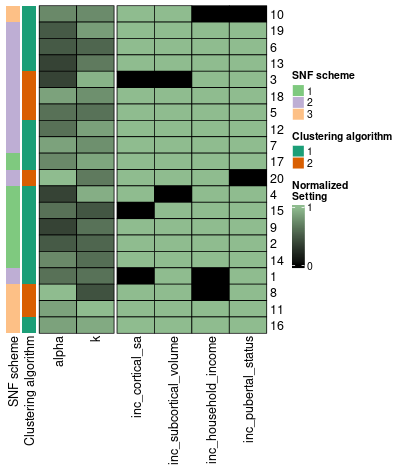}
    \end{minipage}
    \hspace{-4.5em} % Adjust the negative horizontal space as needed
    \begin{minipage}{0.5\textwidth}
        \centering
        \includegraphics[width=0.82\textwidth]{ari_mc_hm_2.png}
    \end{minipage}
    \caption{Side-by-side heatmaps of the settings data frame and corresponding adjusted Rand indices of generated cluster solutions (Figure~\ref{fig:ari_mc_hm}), partitioned by meta cluster.}
    \label{fig:sm_heatmap}
\end{figure}

Plotting settings in the SNF config alongside the meta cluster heatmap can make it easier to identify which of those settings may have been responsible for the overall meta clustering structure observed.
The \code{config_heatmap()} function converts all the columns of the settings data frame (and/or weights matrix, controlled by the function's arguments) to scaled numbers between 0 and 1, then plots those values sorted by the same ordering used for meta cluster identification.
Note that \code{config_heatmap()} is a descriptive alias for \code{plot.snf_config()}.
For example, Figure~\ref{fig:sm_heatmap} suggests that the one cluster solution in meta cluster A is likely driven by the simultaneous dropout of household income and pubertal status features and that meta cluster D is likely driven by usage of the first ("individual") SNF scheme.
Particularly notable columns can be easily referenced as annotations in the \code{meta_cluster_heatmap()}, as all of these columns are present in the extended solutions data frame.
\begin{CodeChunk}
\begin{CodeInput}
R> config_heatmap(sc_6, order = meta_cluster_order_1, hide_fixed = TRUE)
\end{CodeInput}
\end{CodeChunk}

\section{Measures of quality and stability}

\subsection{Quality measures}

\pkg{metasnf} makes it easy to evaluate common clustering measures of quality and stability.
The code below shows the syntax required to calculate Silhouette scores \citep{rousseeuw_silhouettes_1987}, Dunn indices \citep{dunn_fuzzy_1973}, and Davies-Bouldin indices \citep{davies_cluster_1979}.
These metrics all require the similarity matrices that were used to generate each cluster solution.
By setting the parameter \code{return_sim_mats = TRUE} when running \code{batch_snf()}, the resulting solutions data frame will store those similarity matrices as an attribute.
Should users wish to examine these similarity matrices manually, they are accessible by passing the solutions data frame into the function \code{sim_mats_list()}.
Silhouette scores are calculated through \code{cluster::silhouette()} \citep{maechler_2023}, while Dunn indices and Davies-Bouldin indices are calculated by wrapping around functions from the \pkg{clv} package \citep{nieweglowski_clv}.
Note that in order to run the functions \code{calculate_dunn_indices()} and \code{calculate_db_indices()}, users will need to have installed the \pkg{clv} package either manually from CRAN or by setting \code{dependencies = TRUE} during the initial \pkg{metasnf} installation.
\begin{CodeChunk}
\begin{CodeInput}
R> sol_df_with_sim_mats <- batch_snf(dl_1, sc_6, return_sim_mats = TRUE)
R> silhouette_scores <- calculate_silhouettes(sol_df_with_sim_mats)
R> dunn_indices <- calculate_dunn_indices(sol_df_with_sim_mats)
R> db_indices <- calculate_db_indices(sol_df_with_sim_mats)
\end{CodeInput}
\end{CodeChunk}
The documentation in those packages provide additional information on how to further process and analyze these quality measures.
Links to the documentation are listed below:

\begin{itemize}
    \item Silhouette documentation from the \pkg{cluster} package: \url{https://www.rdocumentation.org/packages/cluster/versions/2.1.4/topics/silhouette}.
    \item Dunn index documentation from the \pkg{clv} package: \url{https://www.rdocumentation.org/packages/clv/versions/0.3-2.1/topics/clv.Dunn}.
    \item Davies-Bouldin documentation from the \pkg{clv} package: \url{https://www.rdocumentation.org/packages/clv/versions/0.3-2.1/topics/clv.Davies.Bouldin}.
\end{itemize}

\subsection{Stability measures}

\pkg{metasnf} offers functionality to evaluate two different measures of stability:

\begin{enumerate}
    \item Pairwise adjusted Rand indices: Information on the overall similarity of cluster solutions generated across resamplings of the data.
    \item Co-clustering fraction: Information on the proportion of times that any pair of observations clustered together across resamplings of the data.
\end{enumerate}

The first step in calculating these stability measures is to use the \code{subsample_dl()} function to generate a list of subsampled versions of a data list.
The number of subsamples examined can be controlled with the \code{n_subsamples} parameter, and the fraction of observations that are sampled (without replacement) in each subsampled data list can be controlled with the \code{subsample_fraction} parameter.
\begin{CodeChunk}
\begin{CodeInput}
R> set.seed(42)
R> dl_ss <- subsample_dl(dl_1, n_subsamples = 50, subsample_fraction = 0.85)
\end{CodeInput}
\end{CodeChunk}
\code{dl_ss} is a list that contains 50 subsamples of the original data list.
Each subsample contains a different random 85\% subset of observations.
Using the function \code{batch_snf_subsamples()}, the SNF pipeline defined in the original config can be applied to each subsampled data list.
Running stability measures on every cluster solution can be computationally quite intensive.
Limiting these analyses to only top candidate solutions, such as those determined after visualization of meta clusters and representative solution filtering, can save a considerable amount of time.
The code below focuses on meta clusters B (the third representative solution) and C (the second representative solution).
Parallel processing can be used to speed up \code{batch_snf_subsamples()} by adjusting the \code{processes} parameter.
\begin{CodeChunk}
\begin{CodeInput}
R> top_sol_df <- rep_sol_df_2[c(2, 3), ]
R> top_config <- attr(top_sol_df, "snf_config")
R> progressr::with_progress({
+    ss_results <- batch_snf_subsamples(dl_ss, top_config)})
\end{CodeInput}
\end{CodeChunk}
\code{batch_snf_subsamples()} returns a list of solutions data frames corresponding to each of the provided data list subsamples.
The function \code{subsample_pairwise_aris()} can then be used to calculate the ARIs between cluster solutions across the subsamples.
Note that the calculation of subsampled cluster solutions below involves unstable spectral clustering that may not be exactly reproducible across all machines (Section~\ref{sec:install}).
\begin{CodeChunk}
\begin{CodeInput}
R> pairwise_aris <- subsample_pairwise_aris(ss_results, verbose = TRUE)
R> names(pairwise_aris)
\end{CodeInput}
\begin{CodeOutput}
"raw_aris"    "ari_summary"
\end{CodeOutput}
\begin{CodeInput}
R> pairwise_aris[["ari_summary"]]
\end{CodeInput}
\begin{CodeOutput}
solution  mean_ari     ari_sd
       1 0.9079004 0.06292643
       2 0.5899691 0.24609176
\end{CodeOutput}
\end{CodeChunk}
\code{pairwise_aris} is a two-item list containing a list of matrices, \code{"raw_aris"}, storing the adjusted Rand indices between subsamples of each cluster solution, and a summary of those matrices, \code{"ari_summary"}, which outlines the average adjusted Rand index across all subsamples for each cluster solution.
The mean ARI value reported for each original cluster solution is calculated by taking the mean of the upper triangle of the corresponding ARI matrix, excluding the diagonal.
Note that the ARI calculated between two cluster solutions from different subsamples only considers those observations that are common to both subsamples.

These ARI matrices can also be visualized directly as heatmaps:
\begin{CodeChunk}
\begin{CodeInput}
R> ComplexHeatmap::Heatmap(
+    pairwise_aris[["raw_aris"]][["s1"]],
+    col = circlize::colorRamp2(c(0, 0.5, 1), c("blue", "white", "red")),
+    heatmap_legend_param = list(color_bar = "continuous",
+      title = "Inter-Subsample\nARI", at = c(0, 0.5, 1)),
+    show_column_names = FALSE, show_row_names = FALSE)
R> ComplexHeatmap::Heatmap(
+    pairwise_aris[["raw_aris"]][["s2"]],
+    col = circlize::colorRamp2(c(0, 0.5, 1), c("blue", "white", "red")),
+    heatmap_legend_param = list(color_bar = "continuous",
+      title = "Inter-Subsample\nARI", at = c(0, 0.5, 1)),
+    show_column_names = FALSE, show_row_names = FALSE)
\end{CodeInput}
\end{CodeChunk}
\begin{figure}[H]
    \centering
    \begin{minipage}{0.5\textwidth}
        \centering
        \includegraphics{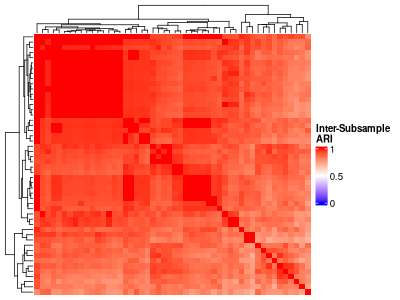}
    \end{minipage}
    \hspace{-4.5em} % Adjust the negative horizontal space as needed
    \begin{minipage}{0.5\textwidth}
        \centering
        \includegraphics{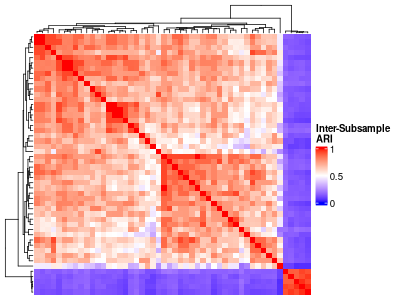}
    \end{minipage}
    \caption{Heatmaps of the adjusted Rand indices from subsampled variations of the representative solutions for meta clusters C (left) and B (right).}
    \label{fig:subsampled_heatmaps}
\end{figure}
Comparing the subsampled adjusted Rand index heatmaps in Figure~\ref{fig:subsampled_heatmaps}, it appears that subsampled variations of meta cluster C remained highly consistent with each other, while subsampled variations of meta cluster B resulted in two distinct sorts of cluster solutions.

\begin{figure}[H]
\centering
\includegraphics[width=0.6\textwidth]{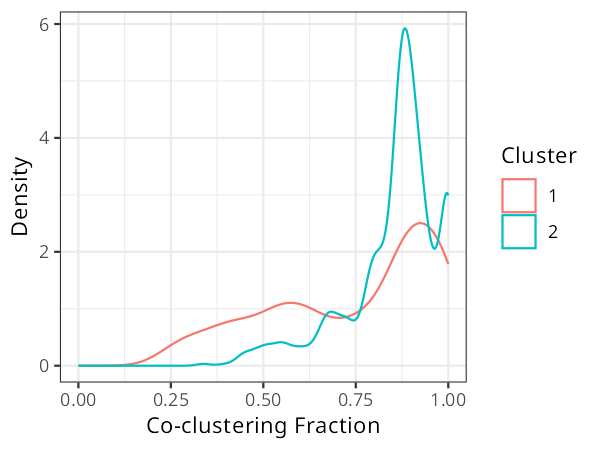}
\caption{Density plot showing the distribution of average co-clustering across all subsamples for each originally co-clustered pair of observations. The red and blue lines show the distribution for pairs of observations that were originally assigned to clusters 1 and 2 respectively.}
\label{fig:cocluster_density}
\end{figure}

\begin{figure}[H]
\centering
\includegraphics{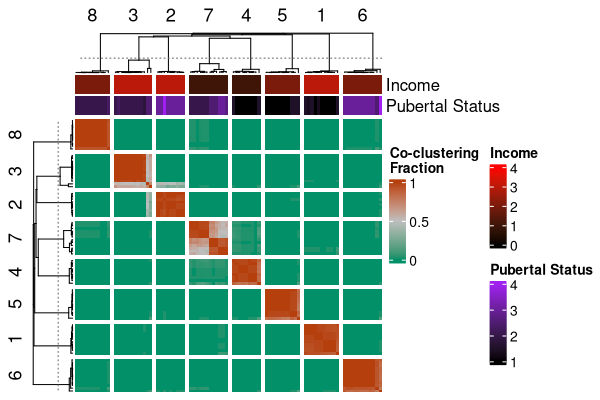}
\caption{Annotated heatmap showing how often all pairs of observations clustered together across subsamples of the data. Cells that are more orange indicate that the corresponding pair of observations clustered together in most subsamples, while cells that are more green indicate that the pair rarely clustered together across subsamples. Segments of the heatmap correspond to which original cluster the corresponding observations belonged to. Within each segment of the heatmap, observations are automatically hierarchically clustered. Annotations at the top of the heatmap show the values of the income and pubertal status features for the corresponding observations.}
\label{fig:cocluster_heatmap}
\end{figure}

The \code{calculate_coclustering} function can be used to calculate how frequently any particular pair of observations cluster together across subsamples of the data.
While the previously calculated inter-subsample ARIs provide some information on the number of cluster solutions associated with a particular dataset-hyperparameter combination, pairwise co-clustering frequencies can provide information on the number of clusters that may best describe that combination. 
\begin{CodeChunk}
\begin{CodeInput}
R> coclustering_results <- calculate_coclustering(ss_results, top_sol_df,
+    verbose = TRUE)
R> names(coclustering_results)
\end{CodeInput}
\begin{CodeOutput}
"cocluster_dfs"     "cocluster_ss_mats" "cocluster_sc_mats"
"cocluster_cf_mats" "cocluster_summary"
\end{CodeOutput}
\end{CodeChunk}
Note that the runtime of the \code{calculate_coclustering} function scales quadratically with the number of observations and linearly with the number of subsamples.
The output of \code{calculate_coclustering} is a list containing the following components:
\begin{enumerate}
    \item \code{cocluster_dfs}: A list of data frames, one per cluster solution, that shows the number of times that every pair of subjects in the original cluster solution occurred in the same subsample, the number of times that every pair clustered together in a subsample, and the corresponding fraction of times that every pair clustered together in a subsample.
    \item \code{cocluster_ss_mats}: The number of times every pair of subjects occurred in the same subsample, formatted as a pairwise matrix.
    \item \code{cocluster_sc_mats}: The number of times every pair of subjects occurred in the same cluster, formatted as a pairwise matrix.
    \item \code{cocluster_cf_mats}: The fraction of times every pair of subjects occurred in the same cluster, formatted as a pairwise matrix.
    \item \code{cocluster_summary}: Among pairs of subjects that clustered together in the original cluster solution, the mean fraction those pairs remained clustered together across the subsample-derived solutions. This information is formatted as a data frame with one row per cluster solution.
\end{enumerate}
As an example, the head of the first data frame in the \code{cocluster_dfs} list looks like:
\begin{CodeChunk}
\begin{CodeInput}
R> cocluster_dfs <- coclustering_results[["cocluster_dfs"]]
R> cocluster_dfs[[1]][1:2, ]
\end{CodeInput}
\begin{CodeOutput}
               obs_1                obs_2 obs_1_clust obs_2_clust
uid_NDAR_INV0567T2Y9 uid_NDAR_INV0J4PYA5F           1           6
uid_NDAR_INV0567T2Y9 uid_NDAR_INV10OMKVLE           1           4
same_solution same_cluster cocluster_frac
           39            0              0
           40            0              0
\end{CodeOutput}
\end{CodeChunk}
The first row indicates that observations \code{NDAR_INV0567T2Y9} and \code{NDAR_INV0J4PYA5F} were originally assigned to clusters 1 and 6 respectively were both present in 39 of the 50 generated subsampled data lists, and were assigned to the same cluster in none of the solutions generated from those subsampled data lists.
As with the \code{ari_summary} values above, note that the \code{cocluster_summary} values shown below may be slightly different for machines with different floating point arithmetic properties than the one used to generate these results.
\begin{CodeChunk}
\begin{CodeInput}
R> coclustering_results[["cocluster_summary"]]
\end{CodeInput}
\begin{CodeOutput}
row avg_cocluster_frac
  1           0.920573
  2           0.816761
\end{CodeOutput}
\end{CodeChunk}
The \code{cocluster_summary} data frame enables quicker comparisons of how stable each original cluster solution was under resampling.
Pairs of observations that clustered together in the first original cluster solution continued to cluster together in around 92\% of the subsamples they were both a part of.
The data generated by \code{calculate_coclustering} can be visualized as either a density plot (Figure~\ref{fig:cocluster_density}) or as a heatmap (Figure~\ref{fig:cocluster_heatmap}) using the \code{cocluster_density} and \code{cocluster_heatmap} functions respectively.
The density plot in \pkg{metasnf} is built using \pkg{ggplot2} \citep{ggplot}.
Both functions require one data frame from the \code{cocluster_dfs} list as input.

For this density example, we examine the second of our two top cluster solutions, as it was a 2-cluster solution resulting in a relatively simple visualization.
\begin{CodeChunk}
\begin{CodeInput}
R> cocluster_density(cocluster_dfs[[2]])
\end{CodeInput}
\end{CodeChunk}
The coclustering density of cluster 2 in Figure~\ref{fig:cocluster_density} shows that most pairs of observations (as indicated by the spike around 0.85) that were originally clustered together in cluster 2 remained clustered together in most of the resampled solutions, a smaller set of pairs of observations (the spike at 1) remained clustered together in every resampled solution, and the remaining pairs of observations remained clustered together in around 40\% to 75\% of the resampled solutions.

With an annotated heatmap, we can investigate if any features in particular were responsible for any observed instability.
We select the first of our two top cluster solutions for this heatmap example, as it most clearly illustrates how a heatmap can assist with this instability inference.
Figure~\ref{fig:cocluster_heatmap} shows that all original clusters were highly robust in this solution, with the exception of cluster 7. 
Two distinct groups of pairs of observations can be seen, indicating that half of the observations assigned to cluster 7 often clustered differently from the other half across resamplings of the data.
The annotations above the segment suggest that this instability was due to half of the observations originally assigned to cluster 7 having had substantially lower pubertal status values than the other half.
\begin{CodeChunk}
\begin{CodeInput}
R> cocluster_heatmap(cocluster_dfs[[1]], dl = dl_1,
+    top_hm = list("Income" = "household_income",
+      "Pubertal Status" = "pubertal_status"),
+    annotation_colours = list(
+      "Pubertal Status" = colour_scale(c(1, 4), min_colour = "black",
+        max_colour = "purple"),
+      "Income" = colour_scale(c(0, 4), min_colour = "black",
+        max_colour = "red")))
\end{CodeInput}
\end{CodeChunk}

\section{Evaluating generalizability through label propagation}

One way to evaluate the generalizability of a cluster solution is by examining how well that cluster solution can describe a new set of observations that were not initially used in the cluster derivation process.
\pkg{metasnf} offers a wrapper around \code{SNFtool::group_predict()} to facilitate label propagation \citep{zhu_2002} of assigned clusters in an extended solutions data frame to a set of new observations.
To propagate cluster labels from a training set to a test set, users should construct an extended solutions data frame containing observations from just the training set and a data list containing observations from both the training and test set.

Prior to the label propagation step, it is recommended that users trim down the extended solutions data frame to just one or a few top candidate solutions.
Evaluating the generalizability of cluster solutions through label propagation is not as well-defined as in the supervised learning case.
Demonstrating that the overall structure of a cluster solution, considered here to be the relative rankings of those clusters across its defining features, persists in held out data can be interpreted as qualitative evidence of the solution's overall generalizability.
Examining how well more than one cluster solution generalizes to the held out data would be analogous to checking more than one supervised learning model against a final test set.
\begin{CodeChunk}
\begin{CodeInput}
R> train_test_split <- train_test_assign(train_frac = 0.8,
+    uids = complete_uids, seed = 42)
R> train_obs <- train_test_split[["train"]]
R> test_obs <- train_test_split[["test"]]
R> train_cort_t <- cort_t[cort_t[["unique_id"]] %in% train_obs, ]
R> train_cort_sa <- cort_sa[cort_sa[["unique_id"]] %in% train_obs, ]
R> train_subc_v <- subc_v[subc_v[["unique_id"]] %in% train_obs, ]
R> train_income <- income[income[["unique_id"]] %in% train_obs, ]
R> train_pubertal <- pubertal[pubertal[["unique_id"]] %in% train_obs, ]
R> train_anxiety <- anxiety[anxiety[["unique_id"]] %in% train_obs, ]
R> train_depress <- depress[depress[["unique_id"]] %in% train_obs, ]
R> test_cort_t <- cort_t[cort_t[["unique_id"]] %in% test_obs, ]
R> test_cort_sa <- cort_sa[cort_sa[["unique_id"]] %in% test_obs, ]
R> test_subc_v <- subc_v[subc_v[["unique_id"]] %in% test_obs, ]
R> test_income <- income[income[["unique_id"]] %in% test_obs, ]
R> test_pubertal <- pubertal[pubertal[["unique_id"]] %in% test_obs, ]
R> test_anxiety <- anxiety[anxiety[["unique_id"]] %in% test_obs, ]
R> test_depress <- depress[depress[["unique_id"]] %in% test_obs, ]
R> train_dl <- data_list(
+    list(train_cort_t, "cort_t", "neuroimaging", "continuous"),
+    list(train_cort_sa, "cortical_sa", "neuroimaging", "continuous"),
+    list(train_subc_v, "subc_v", "neuroimaging", "continuous"),
+    list(train_income, "household_income", "demographics", "continuous"),
+    list(train_pubertal, "pubertal_status", "demographics", "continuous"),
+    uid = "unique_id")
R> train_target_dl <- data_list(
+    list(train_anxiety, "anxiety", "behaviour", "ordinal"),
+    list(train_depress, "depressed", "behaviour", "ordinal"),
+    uid = "unique_id")
R> set.seed(42)
R> sc_7 <- snf_config(train_dl, n_solutions = 5, min_k = 10, max_k = 30)
R> train_sol_df <- batch_snf(train_dl, sc_7)
R> ext_sol_df <- extend_solutions(train_sol_df, train_target_dl)
\end{CodeInput}
\end{CodeChunk}
At this stage, users should ideally reduce their extended solutions data frame to just the top cluster solution(s) that are considered to be worth validating.
The code below shows propagation of cluster labels for only the top cluster solution as determined by having the lowest minimum association p-value across features in the target data list.
\newpage
\begin{CodeChunk}
\begin{CodeInput}
R> lowest_min_pval <- min(ext_sol_df[["min_pval"]])
R> top_df <- ext_sol_df[which(ext_sol_df[["min_pval"]] == lowest_min_pval), ]
R> full_dl <- data_list(
+    list(cort_t, "cort_t", "neuroimaging", "continuous"),
+    list(cort_sa, "cort_sa", "neuroimaging", "continuous"),
+    list(subc_v, "subc_v", "neuroimaging", "continuous"),
+    list(income, "household_income", "demographics", "continuous"),
+    list(pubertal, "pubertal_status", "demographics", "continuous"),
+    uid = "unique_id")
R> propagated_labels <- label_propagate(top_df, full_dl)
R> head(propagated_labels)
\end{CodeInput}
\begin{CodeOutput}
                   uid     group 1
1 uid_NDAR_INV0567T2Y9 clustered 1
2 uid_NDAR_INV0J4PYA5F clustered 2
3 uid_NDAR_INV10OMKVLE clustered 1
4 uid_NDAR_INV15FPCW4O clustered 1
5 uid_NDAR_INV19NB4RJK clustered 1
6 uid_NDAR_INV1HLGR738 clustered 1
\end{CodeOutput}
\end{CodeChunk}

\section{Additional plots}

\subsection{Feature plots}

\begin{figure}[H]
    \centering
    \begin{minipage}{0.3\textwidth}
        \centering
        \includegraphics{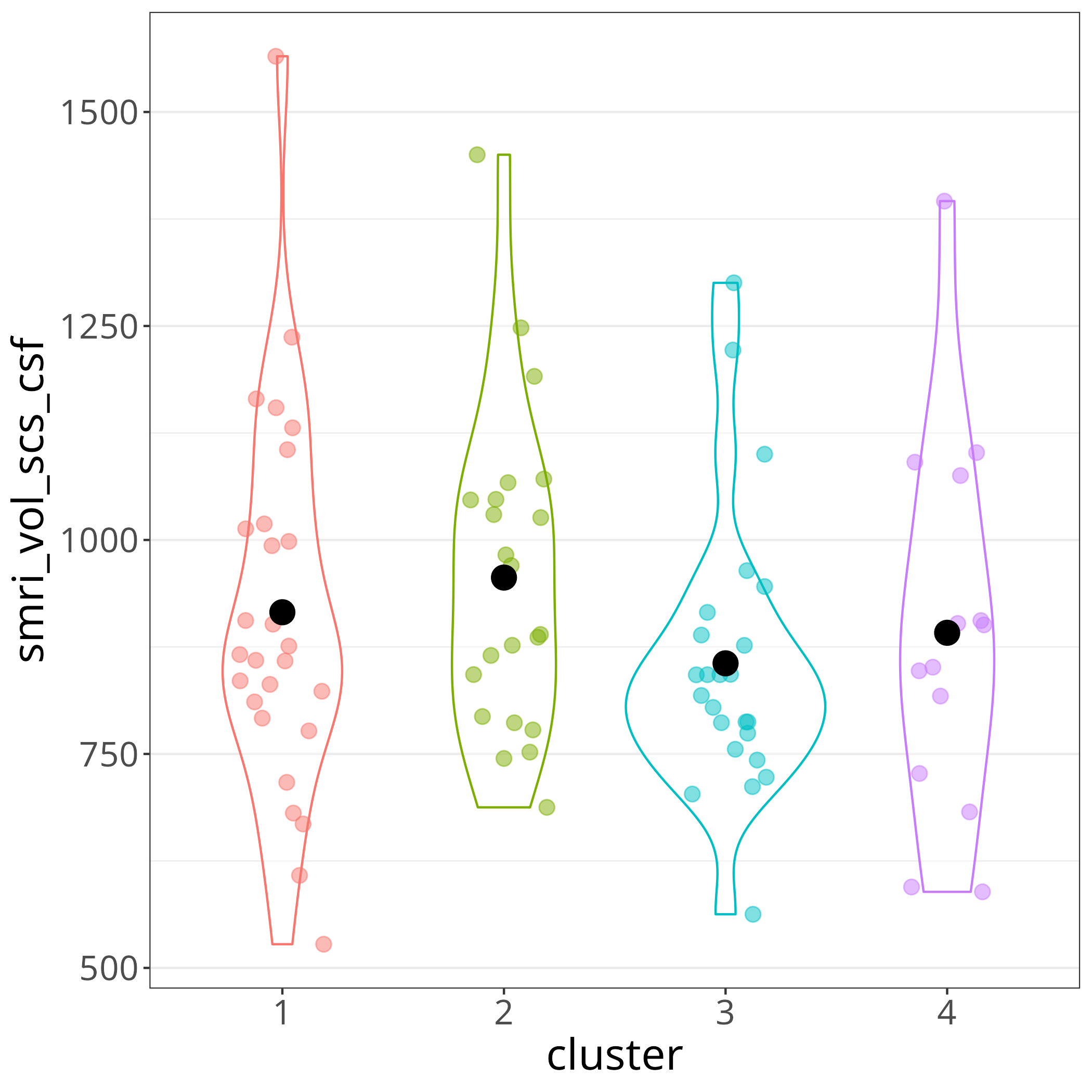}
        \subcaption{}
    \end{minipage}
    \begin{minipage}{0.3\textwidth}
        \centering
        \includegraphics{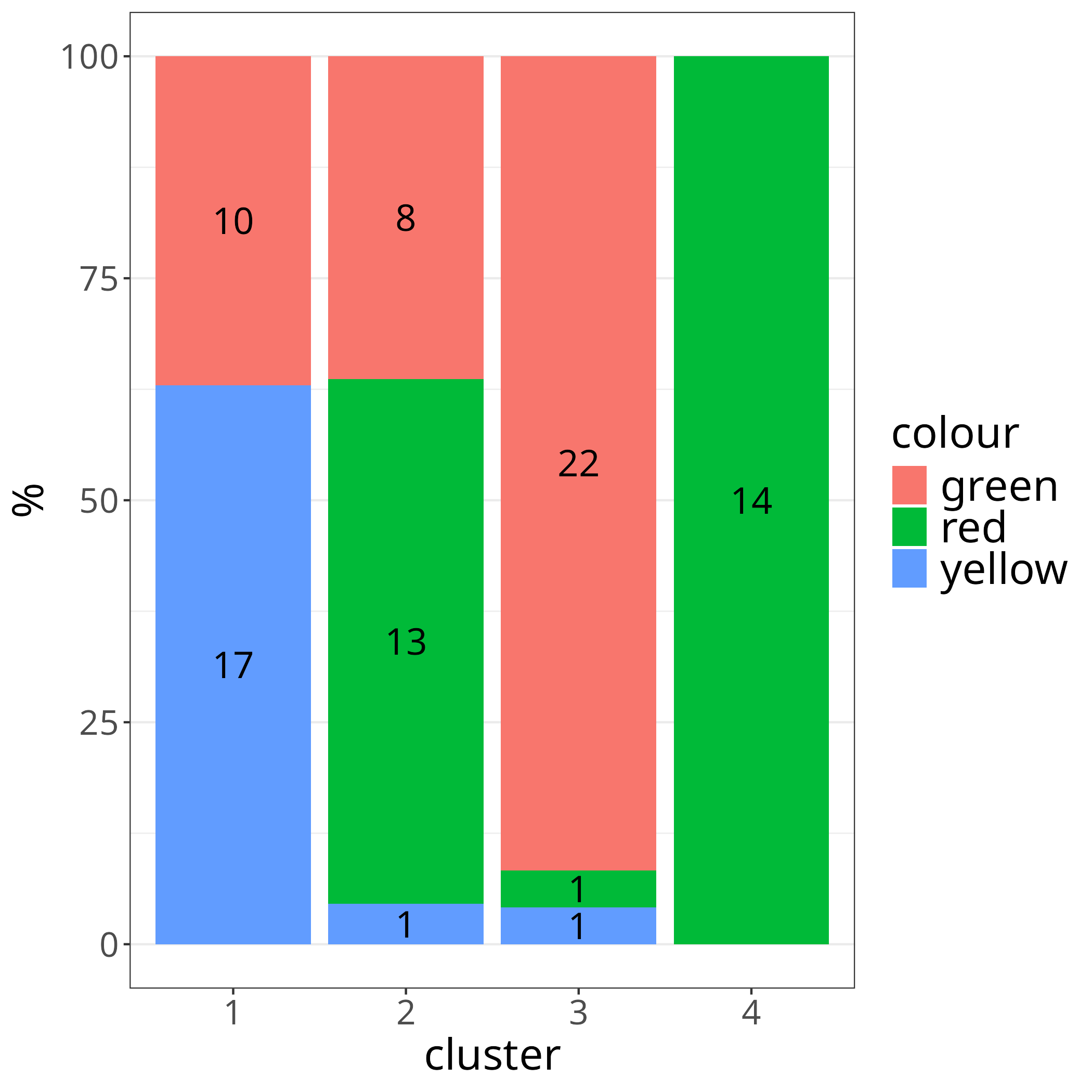}
        \subcaption{}
    \end{minipage}
    \begin{minipage}{0.3\textwidth}
        \centering
        \includegraphics{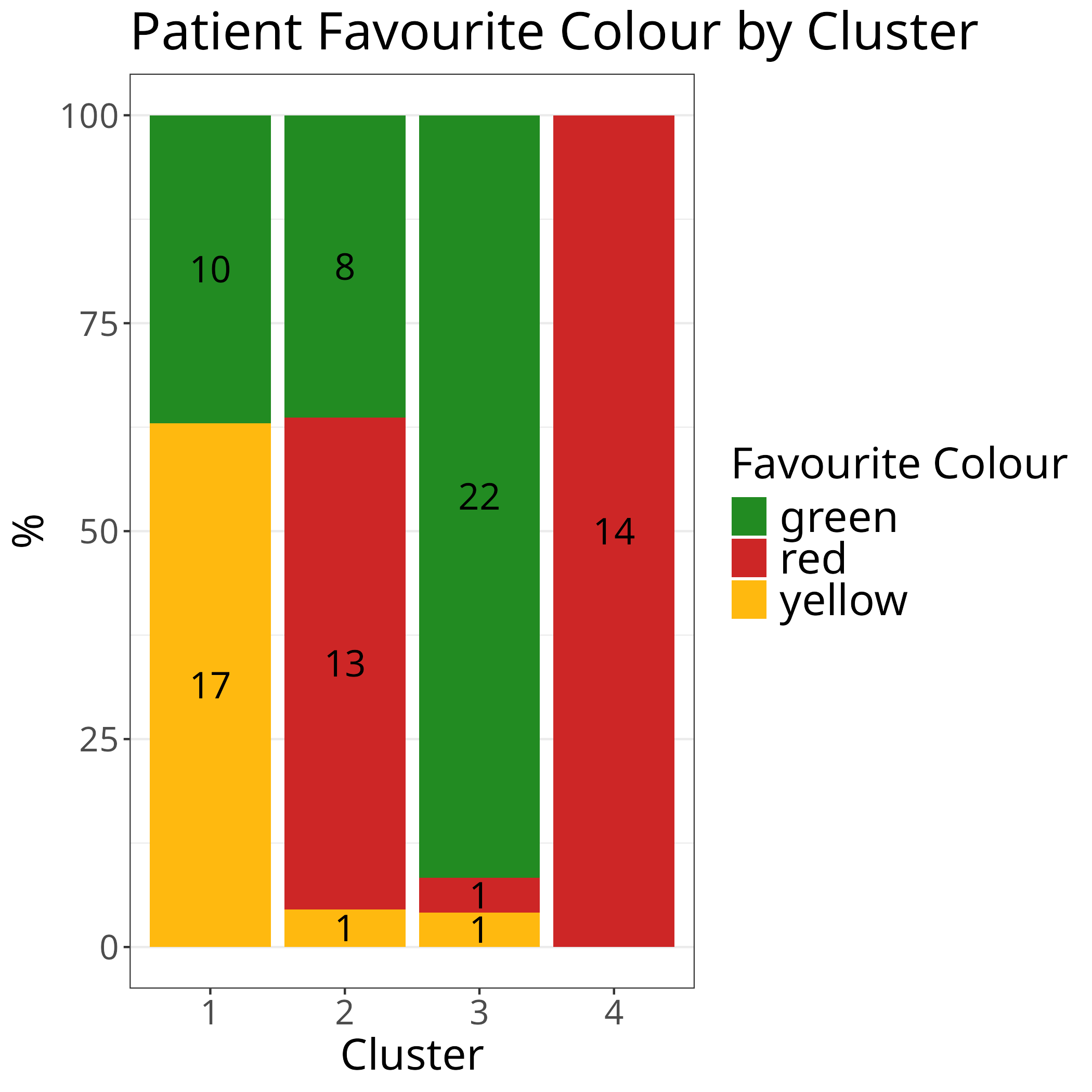}
        \subcaption{}
    \end{minipage}
    \caption{Example plots generated by auto plotting functionality. a) A jitter plot showing the distribution of a neuroimaging measure against cluster membership. b) A bar plot showing the distribution of favourite colour, a categorical feature, by cluster membership. c) A manually revised version of the previous plot.}
    \label{fig:plot_list}
\end{figure}

The \code{autoplot} function allows users to quickly build skeletons of \pkg{ggplot2} \citep{ggplot} plots that show precisely how input and target features are distributed over generated cluster solutions.
\begin{CodeChunk}
\begin{CodeInput}
R> dl_2 <- data_list(
+    list(subc_v, "subcortical_volume", "neuroimaging", "continuous"),
+    list(income, "household_income", "demographics", "continuous"),
+    list(fav_colour, "favourite_colour", "misc", "categorical"),
+    list(pubertal, "pubertal_status", "demographics", "continuous"),
+    list(anxiety, "anxiety", "behaviour", "ordinal"),
+    list(depress, "depressed", "behaviour", "ordinal"),
+    uid = "unique_id")
R> set.seed(42)
R> sc_8 <- snf_config(dl_2, n_solutions = 2, min_k = 20, max_k = 50)
R> sol_df_2 <- batch_snf(dl_2, sc_8)
R> plot_list <- auto_plot(sol_df_row = sol_df_2[1, ], dl = dl_2)
\end{CodeInput}
\end{CodeChunk}
The \code{plot_list} contains plots of each feature in the provided data list organized according to the cluster solution in the provided solutions data frame row.
Features that are non-categorical appear as jitter plots while those that are categorical appear as bar plots.
Any generated plot can easily be modified through normal \pkg{ggplot2} syntax.
Examples of generated plots are shown in Figure~\ref{fig:plot_list}.
\begin{CodeChunk}
\begin{CodeInput}
plot_list[["colour"]]
plot_list[["smri_vol_scs_csf"]]
plot_list[["colour"]] +
  ggplot2::labs(fill = "Favourite Colour", x = "Cluster",
    title = " Favourite Colour by Cluster") +
  ggplot2::scale_fill_manual(values = c("green" = "forestgreen",
    "red" = "firebrick3", "yellow" = "darkgoldenrod1"))
\end{CodeInput}
\end{CodeChunk}

\subsection{Correlation plots}

Examining the correlation structure of the input feature space can provide useful information regarding what sorts of cluster solutions may emerge from the data.
The code below shows how to construct an association p-value matrix, which contains the pairwise association p-values for all features in a data list, as well as how to visualize that matrix as a heatmap.
\begin{CodeChunk}
\begin{CodeInput}
R> cort_sa_minimal <- cort_sa[, 1:5]
R> dl_3 <- data_list(
+    list(cort_sa_minimal, "cortical_sa", "neuroimaging", "continuous"),
+    list(income, "household_income", "demographics", "ordinal"),
+    list(pubertal, "pubertal_status", "demographics", "continuous"),
+    list(fav_colour, "favourite_colour", "demographics", "categorical"),
+    list(anxiety, "anxiety", "behaviour", "ordinal"),
+    list(depress, "depressed", "behaviour", "ordinal"),
+    uid = "unique_id")
R> assoc_pval_matrix <- calc_assoc_pval_matrix(dl_3)
R> assoc_pval_heatmap(assoc_pval_matrix)
\end{CodeInput}
\end{CodeChunk}

\begin{figure}[H]
\centering
\includegraphics{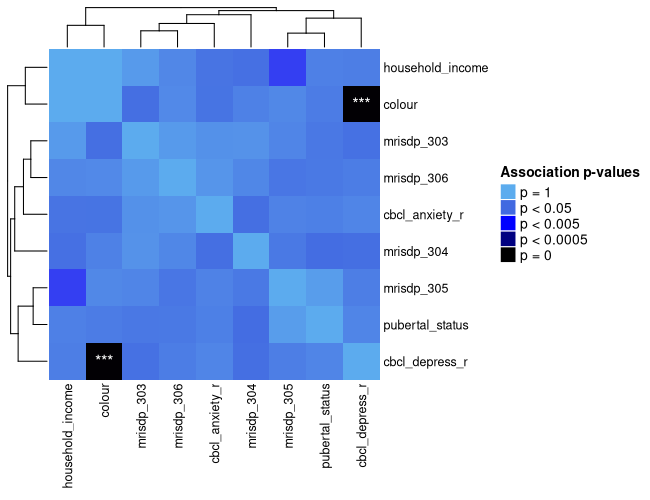}
\caption{
    Heatmap of p-values of pairwise associations between input features in a provided data list.
    Uncorrected significance stars are added by default.
}
\label{fig:ap_hm}
\end{figure}

\begin{figure}[H]
\centering
\includegraphics{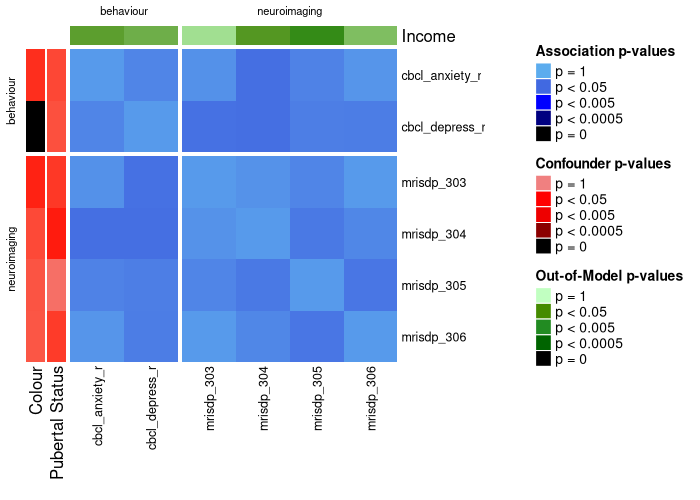}
\caption{
    Heatmap of p-values of pairwise associations between input features in a provided data list.
    Annotations distinguish user-selected features as confounders or out-of-model measures.
    The heatmap is sliced into sections based on feature domains, as specified during data list creation.
}
\label{fig:ap_hm2}
\end{figure}

Features in Figure~\ref{fig:ap_hm} can be highlighted as confounding features or out-of-model measures.
Additionally, the heatmap can be divided into sections based on feature domains (Figure~\ref{fig:ap_hm2}).
\begin{CodeChunk}
\begin{CodeInput}
R> assoc_pval_heatmap(assoc_pval_matrix,
+    confounders = list("Colour" = "colour",
+      "Pubertal Status" = "pubertal_status"),
+    out_of_models = list("Income" = "household_income"), dl = dl_3,
+    split_by_domain = TRUE)
\end{CodeInput}
\end{CodeChunk}

\subsection{Alluvial plots} \label{sec:alluvial}

Alluvial plots can be used in \pkg{metasnf} to visualize how differences in cluster number can change the distribution of observations from the same clustering pipeline.
These plots are created using the package \pkg{ggalluvial} \citep{ggalluvial-package}.
In addition to a typical solutions data frame, alluvial plots require the pre-clustered final similarity matrices as well as a list of clustering functions that control the different numbered cluster solutions to plot along the x-axis.
An example of an alluvial plot is shown below.
More information on clustering algorithms are presented in Appendix~\ref{app:cluster}.
\begin{CodeChunk}
\begin{CodeInput}
R> dl_4 <- data_list(
+    list(expression_df, "genes_1_and_2_exp", "gene_expression",
+      "continuous"),
+    list(methylation_df, "genes_1_and_2_meth", "gene_methylation",
+      "continuous"),
+    list(gender_df, "gender", "demographics", "categorical"),
+    list(diagnosis_df, "diagnosis", "clinical", "categorical"),
+    uid = "patient_id")
R> set.seed(42)
R> sc_9 <- snf_config(dl_4, n_solutions = 1, max_k = 40)
R> sol_df_3 <- batch_snf(dl_4, sc_9, return_sim_mats = TRUE)
R> sim_mats <- sim_mats_list(sol_df_3)
R> similarity_matrix <- sim_mats[[1]]
R> cluster_sequence <- list(spectral_two, spectral_three, spectral_four)
R> alluvial_cluster_plot(cluster_sequence = cluster_sequence,
+    similarity_matrix = similarity_matrix, dl = dl_4,
+    key_outcome = "gender", key_label = "Gender",
+    extra_outcomes = "diagnosis", title = "Gender Across Cluster Counts")
\end{CodeInput}
\end{CodeChunk}

\begin{figure}[H]
\centering
\includegraphics[width=0.6\textwidth]{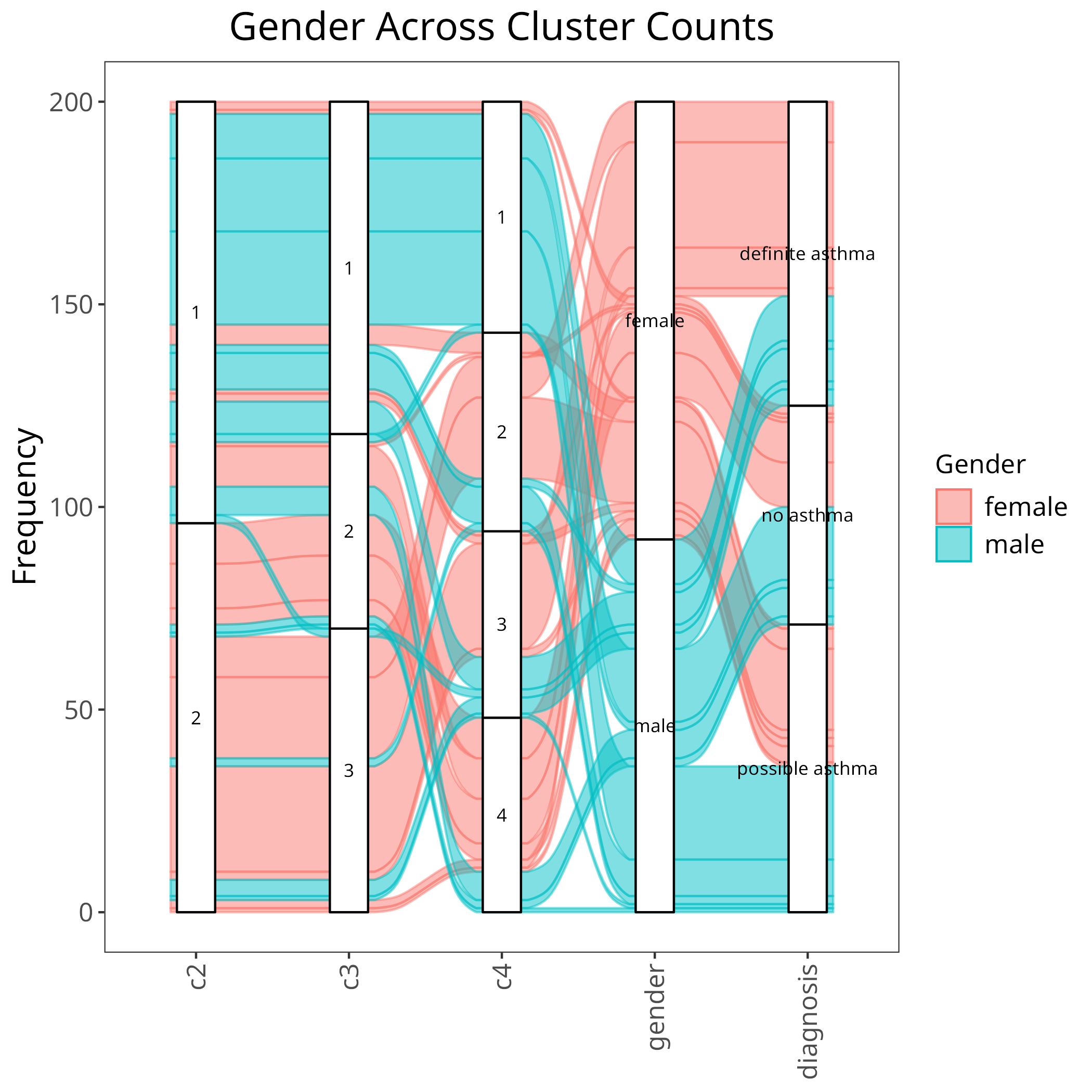}
\caption{
    Alluvial plot showing the distribution of observations across different numbered clustered solutions of a single similarity matrix.
}
\label{fig:alluvial_plot}
\end{figure}

\subsection{Similarity matrix heatmaps}

Similarity matrices are the final product of similarity network fusion.
These matrices are intended to contain the pairwise similarities between all observations based on all the provided input data.
These matrices are generally treated as intermediate products in \pkg{metasnf}, but can be explicitly returned from \code{batch_snf} with the \code{return_similarity_matrices} parameter.
This section builds off of the variables generated in Appendix~\ref{sec:alluvial}.
\begin{CodeChunk}
\begin{CodeInput}
R> similarity_matrix_heatmap(
+    similarity_matrix = similarity_matrix,
+    cluster_solution = sol_df_3[1, ],
+    heatmap_height = grid::unit(10, "cm"),
+    heatmap_width = grid::unit(10, "cm"))
\end{CodeInput}
\end{CodeChunk}

\begin{figure}[H]
\centering
\includegraphics[width=0.6\textwidth]{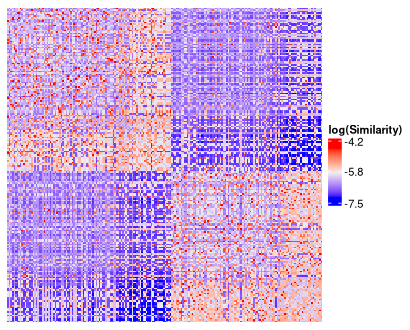}
\caption{
    Plot of a final SNF-derived patient similarity matrix.
}
\label{fig:similarity_matrix_hm}
\end{figure}
The heatmap shown in Figure~\ref{fig:similarity_matrix_hm} can be customized and annotated in the same ways as the meta cluster heatmaps (Figure~\ref{fig:annotated_ari_hm}).

\subsection{Other Manhattan plots}

\begin{figure}[H]
\centering
\includegraphics{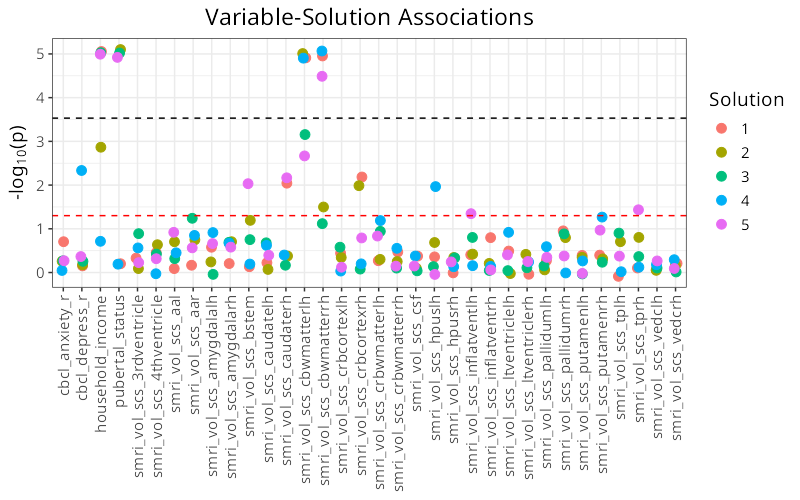}
\caption{
    Example plot generated by \code{esm\_manhattan()}.
    Rows of the extended solutions data frame are provided as inputs and feature separation for each solution is plotted on the same set of axes.
    Solutions are separated by colour.
    The black and red horizontal lines are placed at the Bonferroni and unadjusted equivalents of $p = 0.05$ respectively.
}
\label{fig:esm_manhattan}
\end{figure}

\begin{figure}[H]
\centering
\includegraphics{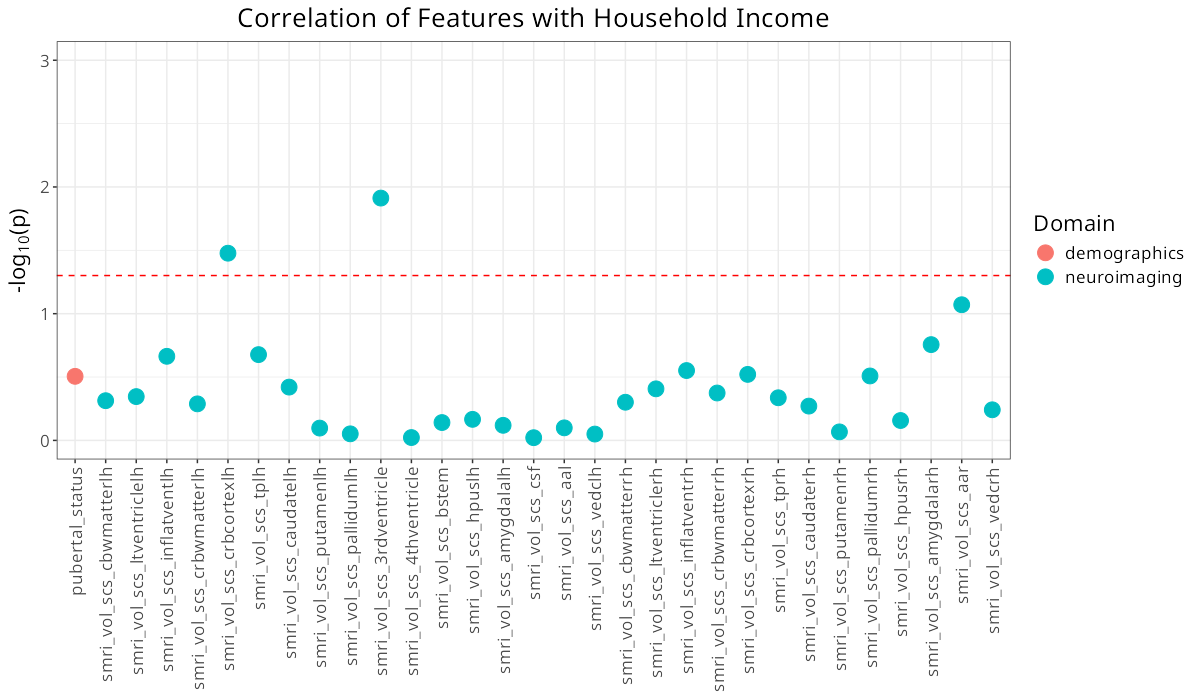}
\caption{
    Example plot generated by \code{var\_manhattan()}.
    One primary feature is specified and the association of that feature with all other features in a provided data list are plotted.
    The red horizontal line is placed at the unadjusted equivalent of $p = 0.05$.
}
\label{fig:var_manhattan}
\end{figure}

Section~\ref{sec:manhattan} introduced \code{mc_manhattan_plot}, which shows feature separation across representative meta clusters.
Two other provided Manhattan plot functions are \code{esm_manhattan_plot} (Figure~\ref{fig:esm_manhattan}), which takes any number of rows from an extended solutions data frame and plots feature separation for all solutions on the same set of axes, and \code{var_manhattan_plot} (Figure~\ref{fig:var_manhattan}), which plots the association of one feature relative to all other features.
\begin{CodeChunk}
\begin{CodeInput}
R> full_dl_2 <- data_list(
+    list(subc_v, "subcortical_volume", "neuroimaging", "continuous"),
+    list(income, "household_income", "demographics", "continuous"),
+    list(pubertal, "pubertal_status", "demographics", "continuous"),
+    list(anxiety, "anxiety", "behaviour", "ordinal"),
+    list(depress, "depressed", "behaviour", "ordinal"),
+    uid = "unique_id")
R> dl_4 <- full_dl_2[1:3]
R> target_dl_2 <- full_dl_2[4:5]
R> set.seed(42)
R> sc_10 <- snf_config(dl_4, n_solutions = 20, min_k = 20, max_k = 50)
R> sol_df_3 <- batch_snf(dl_4, sc_10)
R> ext_sol_df_3 <- extend_solutions(sol_df_3, target_dl_2, dl_4,
+    min_pval = 1e-10)
R> esm_manhattan_plot(ext_sol_df_3[1:5, ],
+    neg_log_pval_thresh = 5, threshold = 0.05, point_size = 3,
+    jitter_width = 0.1, jitter_height = 0.1,
+    plot_title = "Feature-Solution Associations", text_size = 14,
+    bonferroni_line = TRUE)
R> var_manhattan_plot(dl_4, key_var = "household_income",
+    plot_title = "Correlation of Features with Household Income",
+    text_size = 16, neg_log_pval_thresh = 3, threshold = 0.05)
\end{CodeInput}
\end{CodeChunk}

%% -- Summary/conclusions/discussion -------------------------------------------

\section{Limitations} \label{sec:limitations}

\pkg{metasnf} may not always be able to identify useful cluster solutions for users.
Datasets that are sufficiently noisy may result in uniformly low quality solutions regardless of hyperparameter variation.
Feature spaces that are too small to generate meaningfully varied hyperparameters from may not result in a diverse set of meta clusters to select from.
Additionally, a dominating source of variance shared among the bulk of features within a dataset may result in producing a nearly homogeneous space of cluster solutions that are all based on that source.
This particular challenge can be partially mitigated by first regressing out that source of variance as discussed in Appendix~\ref{sec:adjust}, but may be impossible to resolve completely in datasets with non-continuous, non-normally distributed features.

\section{Conclusion} \label{sec:conclusion}

This paper introduces the \proglang{R} package \pkg{metasnf}, which contains a suite of functions facilitating researchers to apply the meta clustering paradigm to SNF-based workflows.
To our knowledge, this is the first software package both within and outside of \proglang{R} designed to assist users with meta clustering, as defined in \cite{caruana_meta_2006}.
As well, this is the first software package to facilitate highly scalable SNF-based cluster analysis.

The meta clustering paradigm enables efficient, data-driven characterization of cluster solutions for complex and noisy datasets.
Through meta clustering, users can get a better understanding of what sorts of distinct cluster solutions can  describe their data and may be able to find a cluster solution better suited to their purpose than would be possible by manual exploration of sensible clustering settings.
By integrating meta clustering with SNF, the benefits of meta clustering are easily extended to multi-modal datasets common to problems such as deriving clinical subtypes from complex patient data.

Apart from this article, documentation for \pkg{metasnf} can also be viewed in the form of \proglang{R} vignettes at \url{https://branchlab.github.io/metasnf/}.

%% -- Optional special unnumbered sections -------------------------------------

\section*{Computational details}

The results in this paper were obtained using \proglang{R} version 4.5.1 and \pkg{metasnf} version 2.1.2 on Fedora Linux 42 with an 11th generation Intel i5-1135G7 central processing unit.
Results were generated on a machine using FlexiBLAS OpenBLAS-OpenMP and LAPACK version 3.12.0.

%% -- Acknowledgements -------------------------------------

\section*{Acknowledgements}
This project was supported by a Precision Child and Youth Mental Health Research Grant from the Hospital for Sick Children.
PSV was supported by a Canadian Institutes of Health Research (CIHR) Canada Graduate Research Scholarship - Doctoral award.
BJC, PAB, AT were supported by funding from CIHR, the Natural Sciences and Engineering Research Council of Canada (NSERC), the Preeclampsia Foundation, and the Peter J Papas award. 
DS was supported by a CIHR Canada Graduate Research Scholarship - Masters award.
SHA currently receives funding from CIHR, the Canada Research Chairs Program, University of Toronto, and the Center for Addiction and Mental Health (CAMH) Foundation.
ALW currently receives funding from CIHR, NSERC, Hospital for Sick Children and Brain Canada.

%% -- Bibliography -------------------------------------------------------------

\bibliography{refs}

%% -- Appendix (if any) --------------------------------------------------------
\newpage

\begin{appendix}

\section{Default S3 plot methods} \label{app:plot}

The following S3 generic \code{plot} methods are available:

\code{plot.ari_matrix()}: Aliased by \code{meta_cluster_heatmap()}, as shown in Figure~\ref{fig:ari_hm}.

\begin{figure}[H]
\centering
\includegraphics{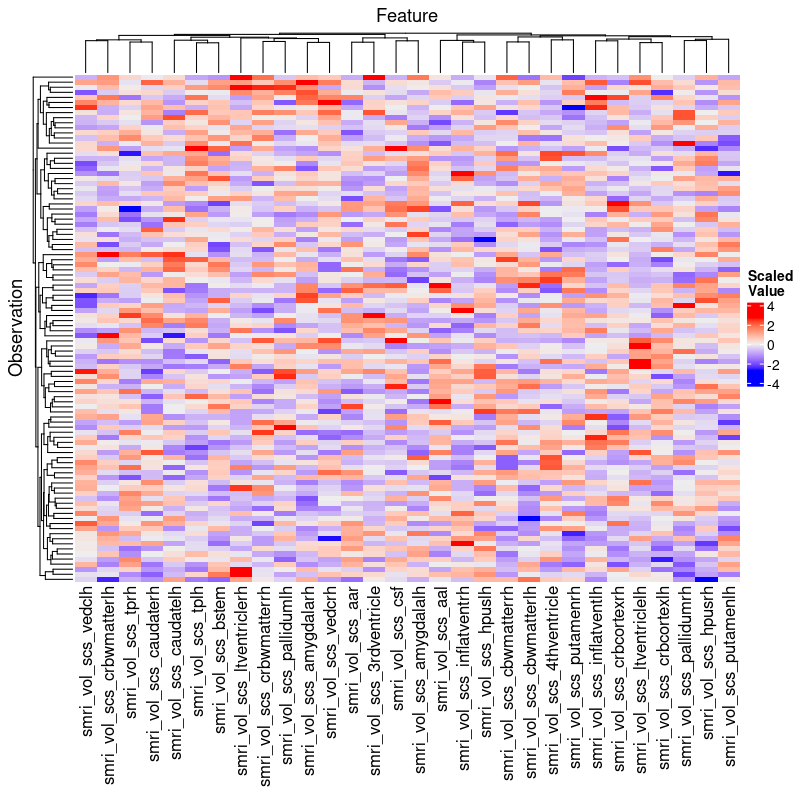}
\caption{
    Heatmap of data distribution within first component of a data list.
    Plot generated by call to \code{plot.data\_list()}.
}
\label{fig:plot_data_list}
\end{figure}

\code{plot.data_list()}: A scaled heatmap of all features contained in the data list, as demonstrated in Figure~\ref{fig:plot_data_list}.
\begin{CodeChunk}
\begin{CodeInput}
R> plot(mock_data_list[1])
\end{CodeInput}
\end{CodeChunk}
\code{plot.solutions_df()}, \code{plot.ext_solutions_df}, \code{plot.t_solutions_df}, and \\ \code{plot.t_ext_solutions_df}, which all plot categorical heatmaps displaying cluster allocations of all observations across all solutions.
\begin{CodeChunk}
\begin{CodeInput}
R> plot(mock_solutions_df, show_column_names = FALSE, show_row_names = TRUE,
+    cluster_rows = FALSE)
\end{CodeInput}
\end{CodeChunk}
\begin{figure}[H]
\centering
\includegraphics{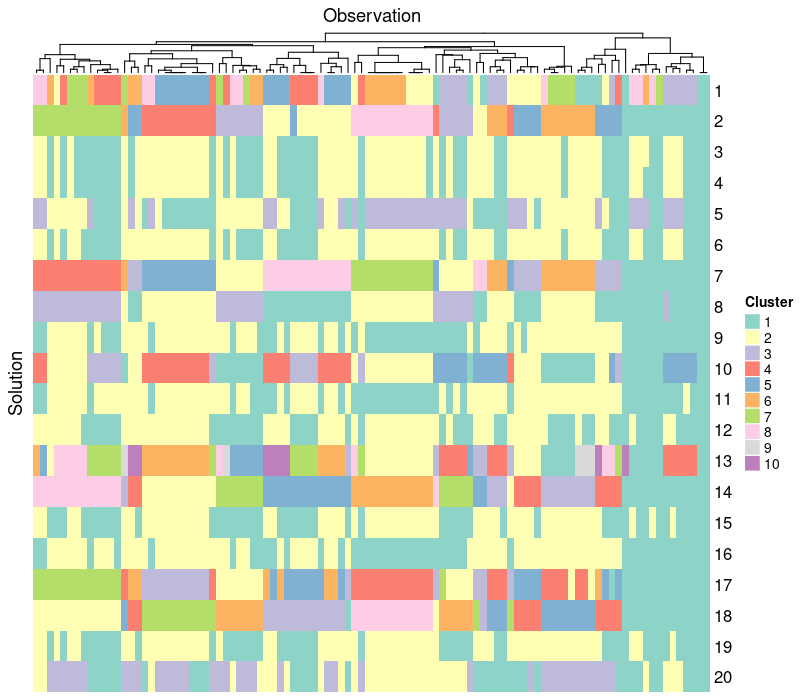}
\caption{
    Heatmap of cluster allocation distribution within a solutions data frame.
    Plot generated by call to \code{plot.solutions\_df()}.
}
\label{fig:sol_df_plot}
\end{figure}
Finally, \code{plot.snf_config()}, \code{plot.settings_df()}, and \code{plot.weights_matrix()} are all aliased by \code{config_heatmap()}, as shown in Figure~\ref{fig:sm_heatmap}.

\section{Additional S3 class methods} \label{app:s3}

In addition to \code{plot()} (Apendix~\ref{app:plot}), a variety of S3 class methods are provided in \pkg{metasnf}.
An overview of many of the available methods are shown in Table~\ref{tab:s3_methods}.

\begin{table}[!h]
\centering
\resizebox{\textwidth}{!}{%
\begin{tabular}{|l|c|c|c|c|c|c|c|c|c|c|c|}
\hline
Class                 & \code{c} & \code{[<-}& \code{[} & \code{print} & \code{summary} & \code{str} & \code{merge} & \code{rbind} & \code{plot} & \code{t} \\
\hline
\code{ari\_matrix}           & \cx        & \cm         & \cm        & \cm            & \cm              & \cm          & \cx            & \cx            & \cm           & \cm        \\
\code{clust\_fns\_list}      & \cm        & \cm         & \cm        & \cm            & \cm              & \cm          & \cm            & \cx            & \cx           & \cx        \\
\code{data\_list}            & \cm        & \cm         & \cm        & \cm            & \cm              & \cm          & \cm            & \cx            & \cm           & \cx        \\
\code{dist\_fns\_list}       & \cx        & \cm         & \cm        & \cm            & \cm              & \cm          & \cm            & \cx            & \cx           & \cx        \\
\code{ext\_solutions\_df}    & \cx        & \cm         & \cm        & \cm            & \cm              & \cm          & \cx            & \cm            & \cm           & \cm        \\
\code{settings\_df}          & \cx        & \cm         & \cm        & \cm            & \cm              & \cm          & \cx            & \cm*           & \cm           & \cm        \\
\code{sim\_mats\_list}       & \cm        & \cm         & \cm        & \cm            & \cm              & \cm          & \cm            & \cx            & \cx           & \cx        \\
\code{snf\_config}           & \cx        & \cm         & \cm        & \cm            & \cm              & \cm          & \cm            & \cx            & \cm           & \cx        \\
\code{solutions\_df}         & \cx        & \cm         & \cm        & \cm            & \cm              & \cm          & \cx            & \cm            & \cm           & \cm        \\
\code{t\_ext\_solutions\_df} & \cx        & \cm         & \cm*       & \cm            & \cm              & \cm          & \cx            & \cm            & \cm           & \cm        \\
\code{t\_solutions\_df}      & \cx        & \cm         & \cm*       & \cm            & \cm              & \cm          & \cx            & \cm            & \cm           & \cm        \\
\code{weights\_matrix}       & \cx        & \cm         & \cm*       & \cm            & \cm              & \cm          & \cx            & \cm*           & \cm           & \cm        \\
\hline
\end{tabular}
}
\caption{
    S3 methods availability for metasnf classes.
    \cm: method is available; \texttimes: the method is explicitly unavailable and signals an error or warning to users; *: the method relies on the base S3 default.
    This table does not include the type-coercing functions such as \texttt{as.data.frame} or \texttt{as.list}.
}
\label{tab:s3_methods}
\end{table}

In addition to the methods listed in Table~\ref{tab:s3_methods}, the following type-coercing functions are available: \code{as.list()} for \code{clust_fns_list}, \code{data_list}, \code{dist_fns_list}, \code{sim_mats_list}, and \\ \code{snf_config};
\code{as.matrix()} for \code{ari_matrix} and \code{weights_matrix}; \code{as.data.frame()} for \\ \code{data_list}, \code{ext_solutions_df}, \code{solutions_df}, \code{t_ext_solutions_df}, \code{t_solutions_df}, and \\ \code{weights_matrix}.

\newpage

\section{Computational runtime and memory scaling} \label{app:runtimes}

\begin{figure}[H]
\centering
\includegraphics{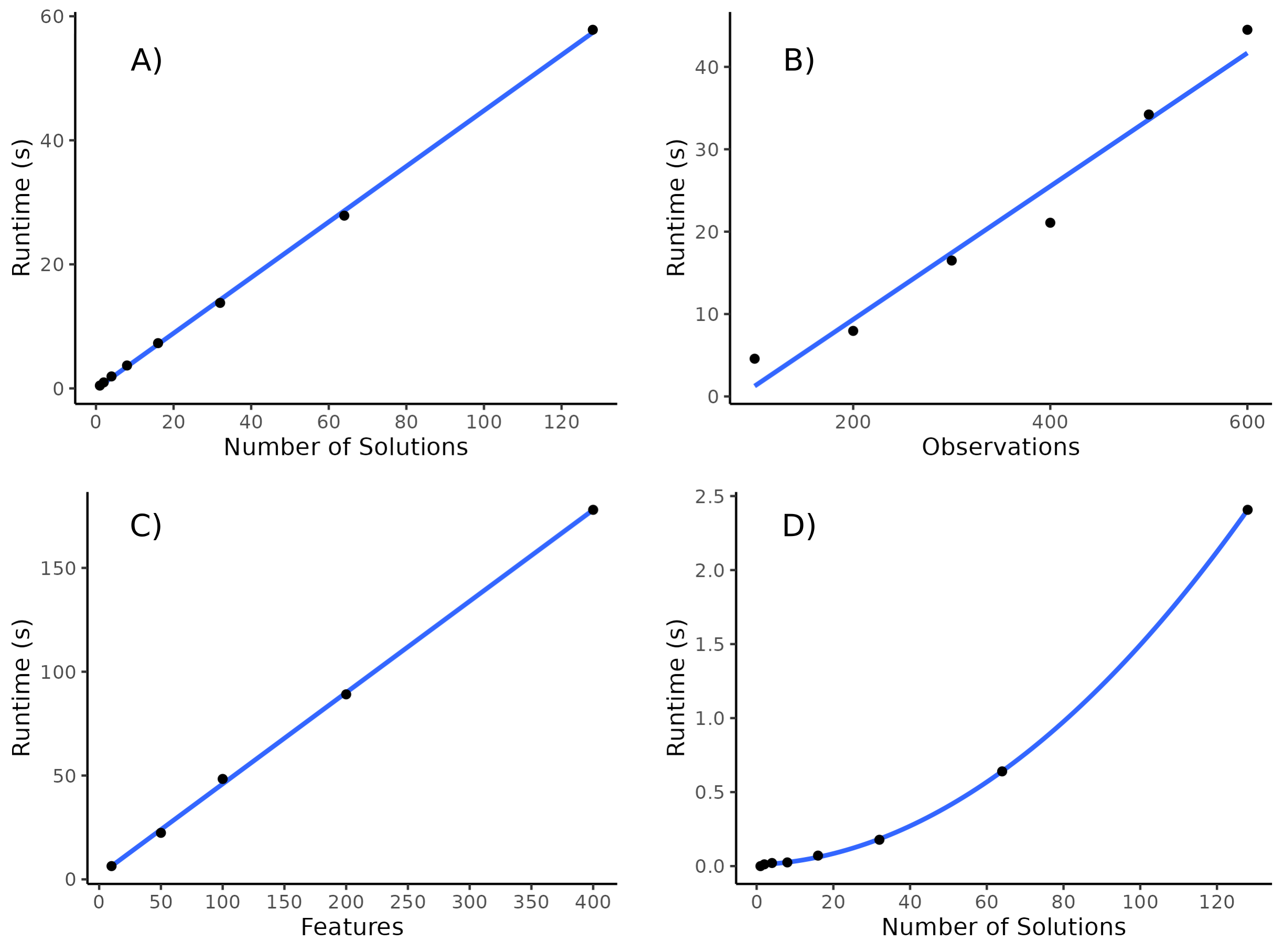}
\caption{Plots A) to C) show how \code{batch\_snf()} runtime scales as major parameters are varied. Plot D) shows how \code{calc\_aris()} runtime scales with the number of cluster solutions.}
\label{fig:runtimes}
\end{figure}

Figure~\ref{fig:runtimes} and Table~\ref{tab:memory} show runtime and memory usage for major operations in \pkg{metasnf}.
Peak memory consumption was monitored using the \pkg{bench} package \citep{bench}.
The code used to generate this data is not shown here for brevity but available in the supplementary replication script.

Users can expect their overall computation time for generating cluster solutions by the call to \code{batch_snf()} to scale linearly with increases in the number of requested cluster solutions, observations, or features.
The call to \code{calc_aris()} that yields the pairwise ARI matrix scales quadratically with the number of solutions included for meta clustering.
Parallelization for these functions as well as \code{extend_solutions()} can be enabled through the \code{processes} parameter.
While these functions are theoretically perfectly parallelizable, details specific to the user's system will determine if the performance improvement offered by parallelization outweighs the overhead required to enable it.

\begin{table}[!h]
\centering
\resizebox{\textwidth}{!}{%
\begin{tabular}{r|r|r|r|r|r|r|r|r}
\hline
Row & Cnt.   & Cat. & p  & n   & Dom. & Sol. & Time (s)  & Peak RAM (MiB)\\
\hline
1 & 0      & 10   & 10 & 100 & 2    & 1    & 1.4 (0.2) & 62.4 (13.3)\\
\hline
2 & 5      & 5    & 10 & 100 & 2    & 1    & 1.4 (0.2) & 62.3 (13.5)\\
\hline
3 & 10     & 0    & 10 & 100 & 2    & 1    & 1.3 (0.2) & 65.5 (12.7)\\
\hline
4 & 10     & 0    & 10 & 100 & 2    & 10   & 13.1 (0.9) & 71.6 (0.2)\\
\hline
5 & 10     & 0    & 10 & 100 & 5    & 1    & 1.4 (0.2) & 71.5  (< 0.1)\\
\hline
6 & 10     & 0    & 10 & 200 & 2    & 1    & 2.6 (0.6) & 72.9  (0.7)\\
\hline
7 & 10     & 0    & 20 & 100 & 2    & 1    & 2.4 (0.3) & 62.7  (14.8)\\
\hline
8 & 10     & 5    & 20 & 400 & 2    & 1    & 12.8 (3.8) & 164.6 (36.8)\\
\hline
9 & 10     & 5    & 20 & 500 & 2    & 1    & 14.2 (5.4) & 195.1 (66.8)\\
\hline
10 & 20     & 0    & 1  & 100 & 2    & 1    & 0.8 (0.3) & 65.4  (12.9)\\
\hline
\end{tabular}
}
\caption{Average clock time and peak memory usage for single-threaded calls of the \code{batch\_snf()} function under different conditions. Averages are calculated from 15 trials evenly distributed across the 3 possible SNF schemes. Cnt. = number of continuous data frames, Cat. = number of categorical data frames, p = number of features per data frame, n = number of observations, Dom. = number of domains that the data frames were distributed over, Sol. = number of solutions generated.}
\label{tab:memory}
\end{table}

Table~\ref{tab:memory} demonstrates peak memory usage of \code{batch\_snf()} under a variety of conditions.
Rows 1 to 3 show that the distribution of data frames across data types does not meaningfully affect runtime or peak memory usage.
The greatest increases in peak memory usage are expected to occur either by integrating several large data frames simultaneously or by generating an extremely large final solutions data frame.
Users can expect parallelization to upscale peak RAM usage and downscale runtime by a factor approximately equal to the number of parallel threads.

\section{Alternative formats for data list generation} \label{app:datalist}

\subsection{Named nested components}

Explicitly specifying each nested component during data list creation can improve code legibility.
\begin{CodeChunk}
\begin{CodeInput}
R> heart_rate_df <- data.frame(unique_id = c("1", "2", "3"),
+    var1 = c(0.04, 0.1, 0.3), var2 = c(30, 2, 0.3))
R> personality_test_df <- data.frame(unique_id = c("1", "2", "3"),
+    var3 = c(900, 1990, 373), var4 = c(509, 2209, 83))
R> dl_5 <- data_list(
+    list(heart_rate_df, "heart_rate", "clinical", "continuous"),
+    list(personality_test_df, "personality_test", "surveys", "continuous"),
+    uid = "unique_id")
\end{CodeInput}
\end{CodeChunk}

\subsection{List of lists}

Instead of specifying each data list component manually, \code{data_list()} can accept a single list of components.
The list-of-lists approach enables users to build their data lists more programmatically.
\begin{CodeChunk}
\begin{CodeInput}
R> list_of_lists <- list(
+    list(heart_rate_df, "data1", "domain1", "continuous"),
+    list(personality_test_df, "data2", "domain2", "continuous"))
R> dl_6 <- data_list(list_of_lists, uid = "unique_id")
\end{CodeInput}
\end{CodeChunk}

\section{Clustering algorithms} \label{app:cluster}

\subsection{The default clustering algorithms list}

SNF produces a single similarity matrix that is meant to describe how similar all the initial observations are to each other across all the provided input features.
Dividing that similarity matrix into subtypes requires can be done using a variety of clustering algorithms.
Within the \pkg{metasnf} package, the clustering function thats used to convert each similarity matrix into a final cluster solution is stored within the \code{snf_config}.
Specifically, the \code{snf_config} contains a list of clustering functions called the clustering functions list (class \code{clust_fns_list}).
By default, this list contains two clustering functions: \code{spectral_eigen()} and \code{spectral_rot()}, which correspond to the spectral clustering method where the number of clusters to partition the matrix into is determined by the eigen-gap or rotation cost heuristics respectively.
\begin{CodeChunk}
\begin{CodeInput}
R> sc_1[["clust_fns_list"]]
\end{CodeInput}
\begin{CodeOutput}
[1] spectral_eigen
[2] spectral_rot
\end{CodeOutput}
\begin{CodeInput}
R> sc_1[["clust_fns_list"]][[1]]
\end{CodeInput}
\begin{CodeOutput}
function (similarity_matrix)
{
    estimated_n <- estimate_nclust_given_graph(W = similarity_matrix,
        NUMC = 2:10)
    nclust_estimate <- estimated_n[["Eigen-gap best"]]
    solution <- SNFtool::spectralClustering(similarity_matrix,
        nclust_estimate)
    return(solution)
}
\end{CodeOutput}
\end{CodeChunk}
Each item of the \code{clust_fns_list} is a function that accepts only one argument, a pairwise similarity matrix, and returns only one value, a vector of cluster assignments for the observations in that similarity matrix.
Notably, the number of clusters is not an argument to the function; these functions must determine the number of clusters to partition a similarity matrix into by themselves, either by hard-coding that number or by implementing a number of clusters heuristic such as is done in \code{spectral_eigen()} and \code{spectral_rot()}.
When each solution is being generated, the function that is selected from the list is controlled by the value in the \code{settings_df} component of the \code{snf_config} object.
\begin{CodeChunk}
\begin{CodeInput}
R> sc_1[["settings_df"]]
\end{CodeInput}
\begin{CodeOutput}
                        1    2    3    4    5    6    7    8    9   10
SNF hyperparameters:
alpha                 0.5  0.4  0.3  0.3  0.5  0.4  0.7  0.8  0.3  0.6
k                      29   26   44   43   29   26   36   21   29   35
t                      20   20   20   20   20   20   20   20   20   20
SNF scheme:
                        2    1    2    1    2    2    2    3    1    3
Clustering functions:
                        1    1    2    1    2    1    1    2    1    1
Distance functions:
CNT                     1    1    1    1    1    1    1    1    1    1
DSC                     1    1    1    1    1    1    1    1    1    1
ORD                     1    1    1    1    1    1    1    1    1    1
CAT                     1    1    1    1    1    1    1    1    1    1
MIX                     1    1    1    1    1    1    1    1    1    1
Component dropout:
cortical_thickness      1    1    1    1    1    1    1    1    1    1
cortical_sa             0    1    0    1    1    1    1    1    1    1
subcortical_volume      1    1    0    0    1    1    1    1    1    1
household_income        0    1    1    1    1    1    1    0    1    0
pubertal_status         1    1    1    1    1    1    1    1    1    0
\end{CodeOutput}
\end{CodeChunk}
This SNF config uses the first clustering function in the list when generating the second cluster solution and uses the second clustering function in the list when generating solutions 1, 3, 4, and 5.

\subsection{Customizing the clustering algorithms list}

Users can either use pre-defined clustering algorithms offered by \pkg{metasnf} or their own clustering algorithm functions when defining a custom clustering algorithms list.

\subsubsection{Pre-defined algorithms}

Pre-defined clustering algorithms offered by \pkg{metasnf} are spectral clustering variations with different numbers of clusters specified:

\begin{enumerate}
    \item \code{spectral_eigen}: Number of clusters determined by eigen-gap heuristic.
    \item \code{spectral_rot}: Number of clusters determined by rotation cost heuristic.
    \item \code{spectral_two}: Yields a two cluster solution.
    \item \code{spectral_three}: Yields a three cluster solution.
\end{enumerate}

And so on, up to \code{spectral_eight}.

A custom clustering algorithms list can be created by passing a named list of functions into the \code{clust_fns} parameter of the \code{snf_config()} function:
\begin{CodeChunk}
\begin{CodeInput}
R> sc_11 <- snf_config(dl_1, n_solutions = 5,
+    clust_fns = list("s2" = spectral_two, "s5" = spectral_five))
R> sc_11[["clust_fns_list"]]
\end{CodeInput}
\begin{CodeOutput}
[1] s2
[2] s5
\end{CodeOutput}
\end{CodeChunk}
The \code{use_default_clust_fns} parameter determines whether these functions will be appended to or overwrite the default functions.
Setting it explicitly to \code{TRUE} will preserve \code{spectral_eigen} and \code{spectral_rot} as the first two functions.
\begin{CodeChunk}
\begin{CodeInput}
R> sc_12 <- snf_config(dl_1, n_solutions = 5,
+    clust_fns = list("s2" = spectral_two, "s5" = spectral_five),
+    use_default_clust_fns = TRUE)
R> sc_12[["clust_fns_list"]]
\end{CodeInput}
\begin{CodeOutput}
[1] spectral_eigen
[2] spectral_rot
[3] s2
[4] s5
\end{CodeOutput}
\end{CodeChunk}

\subsubsection{User-defined algorithms}

User-defined functions can be used to construct the clustering functions list.
These functions should adhere to the following format:

\begin{enumerate}
    \item The function takes a single $N \times N$ similarity (not distance) matrix as its only input.
    \item The function returns a numeric vector of cluster assignments as its only output.
\end{enumerate}

To create variation in the number of clusters generated by an algorithm, users should either build into the function a mechanism for varying the number of clusters or should provide a separate functions for each number of clusters to explore.

\subsection{Non-automated clustering}

Some clustering algorithms require manual intervention that cannot easily be automated within a single function.
Users can extract the final fused networks associated with each SNF solution for subsequent manual clustering by setting \code{return_sim_mats = TRUE} during the call to \code{batch_snf()}.

\section{Distance metrics} \label{app:distance}

Like with clustering algorithms, \pkg{metasnf} enables users to customize what distance metrics are used in the SNF pipeline.
All information about distance metrics are stored in a \code{dist_fns_list} object.

By default, \code{batch_snf()} will create its own internal \code{distance_metrics_list} by calling the \code{generate_distance_metrics_list()} function with no additional arguments.
The distance metrics list is a nested list, where the first layer of the list contains 5 lists (one for each feature type) and the second layer of the list contains an arbitrary number of distance functions for those feature types.
The distance calculating functions themselves accept raw input data frames and a vector of feature weights and return a matrix object containing pairwise distances of all observations.
By default, only a single distance metric function is used for each data type: simple Euclidean distance for continuous, discrete, or ordinal data, and Gower's distance for categorical and mixed data.
In the exact same manner as the clustering functions are controlled, which distance function is used when multiple are available for a single feature type will be determined by the values stored in the \code{settings_df} component of the \code{snf_config} object.

\subsection{Using custom distance functions}

Users can customize the distance functions list by supplying pre-defined distance metric functions in \pkg{metasnf} or by writing their own distance metric functions.

\subsubsection{Pre-defined distance metrics functions}

\begin{enumerate}
    \item \code{euclidean_distance}: Simple Euclidean distance.
    \item \code{sn_euclidean_distance}: Standardizes and normalizes data prior to Euclidean distance calculation.
    \item \code{gower_distance}: Gower's distance calculates dissimilarity over mixed-type features by averaging contributions from different types of variables through different distance measures. The \pkg{metasnf} package relies on the Gower's distance implementation provided by the \pkg{cluster} package \citep{maechler_2023}. The precise breakdown of how Gower's distance handles each type of variable can be learned more about in its original publication \cite{gower_general_1971}.
    \item \code{siw_euclidean_distance}: Squared, including weights, Euclidean distance. Apply feature weights if provided to data frame, then calculates Euclidean distance, then squares the results.
    \item \code{sew_euclidean_distance}: Squared, excluding weights, Euclidean distance. Apply square root of feature weights to data frame, then calculates Euclidean distance, then squares the results.
    \item \code{hamming_distance}: Hamming distance. When calculated between two vectors of equal length, the distance is the sum of the number of positions where the two vectors are not identical.
\end{enumerate}

Any of these functions can be accessed upon loading \pkg{metasnf} and can be formatted into a custom \code{dist_fns_list} as follows:
\begin{CodeChunk}
\begin{CodeInput}
R> sc_13 <- snf_config(dl_1, n_solutions = 5,
+    cnt_dist_fns = list("standard_norm_euclidean" = sn_euclidean_distance),
+    dsc_dist_fns = list("standard_norm_euclidean" = sn_euclidean_distance),
+    use_default_dist_fns = TRUE)
R> sc_13[["dist_fns_list"]]
\end{CodeInput}
\begin{CodeOutput}
Continuous (2):
[1] euclidean_distance
[2] standard_norm_euclidean
Discrete (2):
[1] euclidean_distance
[2] standard_norm_euclidean
Ordinal (1):
[1] euclidean_distance
Categorical (1):
[1] gower_distance
Mixed (1):
[1] gower_distance
\end{CodeOutput}
\end{CodeChunk}
Once constructed with the custom functions, the \code{snf_config} object can be used as normal.

\subsubsection{User-defined distance metrics functions}

Users can pass in their own distance metric functions when creating a distance metrics list.
These functions should adhere to the following format:

\begin{enumerate}
    \item The first parameter, \code{df}, is a data frame that contains no UID column and contains an arbitrary number of feature columns.
    \item The second parameter, \code{weights_row} is a numeric vector of weights corresponding to the features in \code{df}.
\end{enumerate}

While it is necessary for the function to accept a \code{weights_row}, it is not necessary for the function to make use of this row.
When writing code to apply weights to features within a function, one common approach is to convert the \code{weights_row} to a diagonal matrix, then matrix multiply \code{df} (as a matrix) with that resulting diagonal weights matrix.
Be cautious of inconsistent default behaviour of \code{base::diag()}; when \code{weights_row} has only a single element (expected for any single-feature data frame), \code{base::diag()} will return a diagonal matrix of ones with the number of rows and columns equal to the value of that single element, rather than the expected behaviour of a $1 \times 1$ matrix containing that single element.
The helper function \code{format_weights_row} or the base \proglang{R} documentation's recommended syntax of \code{diag(x, nrow = length(x))} can be used to avoid this problem.
Source code for any pre-defined distance function can also serve as a template for how weights can be applied.

\section{Association-signal-annotation boosted similarity network fusion} \label{app:absnf}

Association-signal-annotation boosted similarity network fusion (ab-SNF; \cite{ruan_using_2019}) is a variation of traditional SNF that weights features based their association p-value with a primary held-out feature of interest.
The procedure has its own corresponding \proglang{R} package, \pkg{abSNF} \citep{ruan_using_2019}.
Functions related to distance metrics (Appendix~\ref{app:distance}) and feature weights can be leveraged to run ab-SNF within \pkg{metasnf}.

Users should first construct a data list containing all input features for clustering as well as the primary held-out feature of interest.
The function \code{calc_assoc_pval_matrix} generates the relevant set of association p-values (referred to as $p_k$ in ab-SNF).
The row or column of this association p-value matrix that corresponds to the feature of interest should be negative log normalized prior to being used as any rows of the weights matrix that users wish to run ab-SNF with.
For those rows, users should also ensure that the \code{sew_euclidean_distance} function is used as the distance metric for any numeric features.
Those rows should also use the individual (1) SNF scheme should users wish to exactly match a traditional ab-SNF workflow. 
Example code is shown below.
\newpage
\begin{CodeChunk}
\begin{CodeInput}
R> dl_7 <- data_list(
+    list(income, "income", "demographics", "ordinal"),
+    list(pubertal, "pubertal_status", "demographics", "continuous"),
+    list(fav_colour, "favourite_colour", "demographics", "categorical"),
+    list(anxiety, "anxiety", "behaviour", "ordinal"),
+    uid = "unique_id")
R> assoc_pval_matrix <- calc_assoc_pval_matrix(dl_7)
R> colour_pk <- assoc_pval_matrix["colour", ]
R> colour_pk <- colour_pk[!names(colour_pk) == "colour"]
R> feature_weights <- -log10(colour_pk)/sum(-log10(colour_pk))
R> dl_7["favourite_colour"] <- NULL
R> sc_14 <- snf_config(dl_7, n_solutions = 1)
R> sc_14[["weights_matrix"]][1, ] <- feature_weights
R> ab_sol_df <- batch_snf(dl_7, sc_14)
\end{CodeInput}
\end{CodeChunk}

\section{Imputation of missing data as a meta clustering parameter} \label{app:imputation}

The influence of imputation on the generated space of cluster solutions can be examined easily in \pkg{metasnf}.
One approach is to generate separate solutions data frames for each imputed version of the data, then combine those solutions data frames and proceed with the rest of the meta clustering pipeline as usual.

An example of this process is shown below.
Two data lists are prepared in place of what could be distinct data lists from imputed variations of the underlying data frames.
\begin{CodeChunk}
\begin{CodeInput}
R> dl_imp1 <- data_list(
+    list(subc_v, "subcortical_volume", "neuroimaging", "continuous"),
+    list(depress, "depressed", "behaviour", "ordinal"),
+    uid = "unique_id")
R> dl_imp2 <- data_list(
+    list(subc_v, "subcortical_volume", "neuroimaging", "continuous"),
+    list(depress, "depressed", "behaviour", "ordinal"),
+    uid = "unique_id")
R> sc_15 <- snf_config(dl_imp1, n_solutions = 5, min_k = 20, max_k = 50)
R> sol_df_imp1 <- batch_snf(dl_imp1, sc_15)
R> sol_df_imp2 <- batch_snf(dl_imp2, sc_15)
R> sol_df <- rbind(sol_df_imp1, sol_df_imp2, reset_indices = TRUE)
R> sol_aris <- calc_aris(sol_df)
\end{CodeInput}
\end{CodeChunk}

\section{Linearly regressing out unwanted signal} \label{sec:adjust}

Clustering solutions reveal patterns of shared variance across the input feature space.
It is possible for one dominating layer of variance to mask another layer of variance that represents a more useful situation to the context.
For example, consider the problem of clustering different attributes of athletes in a sport to identify distinct styles of playing.
If the data is collected from players at both an amateur and professional setting, the differences between the amateur and professional players may outweigh the differences between the different playing styles. 
The result may be a somewhat uninteresting two-cluster solution that separates players by skill level.
Transforming the input feature space into the residuals of the linear model \code{feature ~ skill_level} will remove the unwanted effect of player skill and make it easier for clustering solutions to pick up on style-based differences.

\pkg{metasnf} offers the \code{linear_adjust()} function for such linear scaling, but be aware that this is only expected to work as intended for continuous features that are normally distributed.
Categorical features cannot be adjusted in this way, while non-normal and non-continuous data may exhibit unusual patterns after adjustment that continue to exhibit non-linear associations with the unwanted signal.

\section{The Cancer Genome Atlas example: glioblastoma multiforme}

\begin{figure}[H]
\centering
\includegraphics{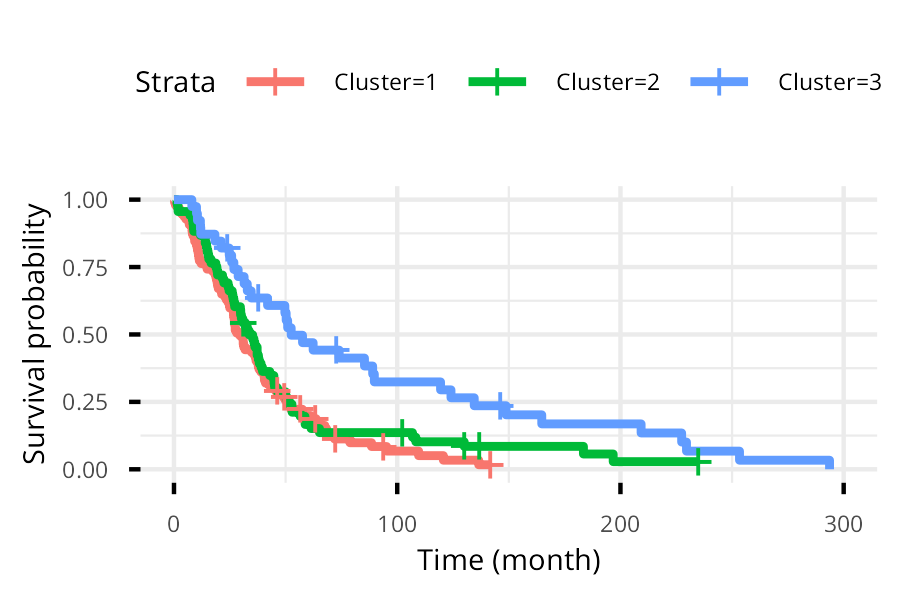}
\caption{Survival probability curve for clustered GBM patients from the TCGA dataset.}
\label{fig:survival}
\end{figure}

Here we apply \pkg{metasnf} to replicate results from real-world glioblastoma multiforme (GBM) data explored in \cite{wang_similarity_2014}.
These results are completely based on freely available data generated by the The Cancer Genome Atlas (TCGA) Research Network: \url{https://www.cancer.gov/tcga}.
Two meta clusters were identified by calculating only 10 solutions from integrated gene expression, methylation, and miRNA expression data of 204 patients with GBM.
One of the meta clusters corresponded to a 3-cluster solution, as was identified in \cite{wang_similarity_2014}.
Subsequent characterization of this meta cluster revealed patients in cluster 3 as having a substantially better survival probability curve compared to those in clusters 1 and 2 (Figure~\ref{fig:survival}).
This survival probability curve as well as the distribution of gene expression and methylation levels for the genes APBA2, GLI2, CTSD, and S100A4 shown in Figure~\ref{fig:gene_hm} closely resemble the clustering results of this dataset presented in the supplementary section of \cite{wang_similarity_2014}.
The code used to generate this data is not shown here for brevity but available in the supplementary replication script.
The survival analysis was conducted using \pkg{survival} \citep{survival-package} and the survival plot was generated using \pkg{survminer} \citep{survminer}.

\begin{figure}[H]
\centering
\includegraphics{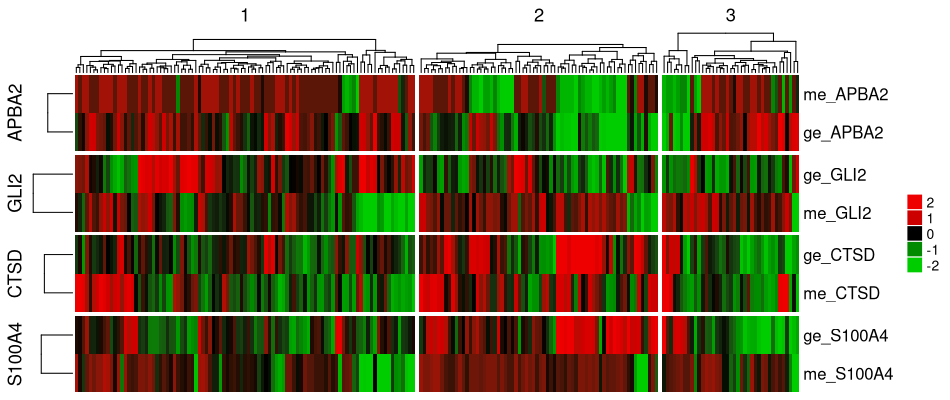}
\caption{He map of gene expression (prefixed with ``ge\_'') and methylation levels (prefixed with ``me\_'') for key genes distinguishing TCGA GBM cluster solution identified in \cite{wang_similarity_2014}.}
\label{fig:gene_hm}
\end{figure}

\end{appendix}

\newpage

\end{document}